\newif\iffigs\figstrue
\documentclass[paper, 12pt, letterpaper]{JHEP}
\def\Bbb{\bf}
\def\C{{\Bbb C}}

\def\Z{{\Bbb Z}}

\def\bearray{\begin{eqnarray}}
\def\eearray{\end{eqnarray}}
\def\bearraynn{\begin{eqnarray*}}
\def\eearraynn{\end{eqnarray*}}
\def\bfig{\begin{figure}}
\def\efig{\end{figure}}

\def\opeq#1{\advance\lineskip#1 \advance\baselineskip#1
        \advance\lineskiplimit#1}

\newtheorem{Proposition}{Proposition}[section]

\newtheorem{Theorem}{Theorem}[section]
\newtheorem{Lemma}{Lemma}[section]
\newtheorem{Corrolary}{Corrolary}[section]

\newcommand{\be}{\begin{equation}}
\newcommand{\ee}{\end{equation}}
\newcommand{\bea}{\begin{eqnarray}}
\newcommand{\eea}{\end{eqnarray}}

\newcommand{\bp}{\begin{Proposition}}
\newcommand{\ep}{\end{Proposition}}
\newcommand{\bt}{\begin{Theorem}}
\newcommand{\et}{\end{Theorem}}
\newcommand{\bl}{\begin{Lemma}}
\newcommand{\el}{\end{Lemma}}
\newcommand{\bc}{\begin{Corrolary}}
\newcommand{\ec}{\end{Corrolary}}
\newcommand{\nn}{\nonumber}

\usepackage{graphics}

\title{String field theory and brane superpotentials}

\author{C.~I.~Lazaroiu\\C.~N.~Yang Institute for Theoretical Physics\\
SUNY at Stony BrookNY11794-3840, U.S.A.\\calin@insti.physics.sunysb.edu}

\abstract{I discuss tree-level amplitudes in cubic topological string field
theory, showing that a certain family of 
gauge conditions leads to an $A_\infty$ algebra of
tree-level string products which define a potential
describing the dynamics of physical states.
Upon using results of modern deformation theory, 
I show that the string moduli space admits two equivalent 
descriptions, one given in standard Maurer-Cartan fashion and
another given in terms of a `homotopy Maurer-Cartan problem',
which describes the critical set of the potential. 
By applying this construction to the topological 
A and B models, I obtain an intrinsic formulation of 
`D-brane superpotentials' in terms of string field theory data. 
This gives a prescription for computing such quantities to all orders, 
and proves 
the equivalence of this formulation with the fundamental 
description in terms of string field moduli. 
In particular, it clarifies the relation between the Chern-Simons/holomorphic 
Chern-Simons actions and the superpotential for A/B-type branes.
}

\begin{document}

\tableofcontents

\pagebreak
\vskip .6in

\section{Introduction}

An important subject in D-brane geometry concerns 
the computation of brane superpotentials, as discussed for the first time in 
 \cite{Douglas_quintic}. In typical examples, one is interested in a D-brane 
wrapping a supersymmetric cycle of a Calabi-Yau threefold, and attempts
to describe its moduli space through the critical set of such a quantity.

It is fair to say that results in this direction have remained somewhat 
imprecise. Part of this lack of precision 
is due to our incomplete understanding of 
the mirror map for open string backgrounds. Another reason 
can be found in the absence of a rigorous 
formulation of the problem. There are at least 
two issues to be addressed before one can gain a
better understanding:

(1)The current definition of `D-brane superpotentials' is based on an
indirect construction involving partially wrapped branes 
which fill the four uncompactified dimensions. Due to standard difficulties 
with flux conservation, this is in fact physically inconsistent unless one 
restricts to non-compact situations (which in themselves have limited physical
relevance) or posits some unspecified 
orientifold constructions which would solve the difficulty. 

(2)A perhaps 
more serious problem is the lack of a precise formulation of the 
relation between flat directions for the superpotential and the 
string theory moduli 
space -- this is currently resolved by {\em assuming} that the 
two spaces coincide, since moduli problems only involve low 
energy dynamics. However, the fundamental description of D-brane 
moduli is through the associated string field theory (see Section 3.4
of this paper), which in our 
case is the string field theory of topological A or B type models 
(the fact that topological string theory suffices 
follows from the results of  \cite{Kachru1}). It is 
not immediately clear how the string field point of view 
relates to the approach advocated in 
 \cite{Douglas_quintic}. 

The purpose of the present note is to initiate a more thorough analysis 
of these issues by addressing the two problems above. 
Our approach is based on a string field theoretic 
point of view, which was advocated in a wider context in 
 \cite{com1, com3} (see also  \cite{Diaconescu} and  \cite{sc, bv}). 
This has the advantage that it provides an {\em intrinsic}
description of  D-brane moduli spaces. As we shall show below, 
string field theory allows for a precise formulation of
brane `superpotentials' in a manner which does not require the introduction 
of partially wrapped branes.  Instead, we shall identify the superpotentials
of  \cite{Douglas_quintic} with a generating function for a collection
of tree-level string amplitudes computed in a certain gauge.
The construction allows us to 
{\em prove} that the string field moduli space 
can be described as the critical set of this function, 
divided by an appropriate group action. 
This serves to clarify the relation between the
two descriptions, and sheds some 
light on the connection of D-brane superpotentials 
to certain mathematical constructions involved in the homological mirror 
symmetry conjecture  \cite{Kontsevich}. 

The note is organized as follows. In Section 2, we explain the construction 
of our potential $W$ as a `generating function' of tree level string 
amplitudes, and show that the coefficients of its 
expansion can be expressed in terms of a collection of 
tree level products  which satisfy the constraints 
of an $A_\infty$ algebra. 
While we shall apply our constructions to the ungraded A/B models
only, the discussion of this (and the next) section is given in 
slightly more abstract terms, and can be applied 
to more general situations, such as the graded string field 
theories of \cite{com3,Diaconescu,sc,bv}. 
In Section 3, we discuss two  
formulations of the brane moduli space, which result by considering the 
string field equations of motion or the critical manifold of the potential, 
and dividing by appropriate symmetries.
While the first description involves the well-known Maurer-Cartan equation
(and the string field theory gauge group ${\cal G}$), the second 
leads to more complicated data, due to the presence of higher order 
terms.
We use the algebraic structure of tree level products to 
give a mathematically precise formulation of the 
second description in terms of 
a homotopy version of the Maurer-Cartan equation and a certain 
effective symmetry algebra. More precisely, we show 
that the critical point condition for $W$ can be expressed in 
terms of a strong homotopy Lie, or $L_\infty$ algebra, the so-called 
{\em commutator algebra} 
of the $A_\infty$ algebra of tree level products. 
The resulting {\em homotopy Maurer-Cartan equation} has a 
symmetry algebra $g_W$ which plays the role of an effective, or `low energy'  
remnant of the string field theory gauge algebra. 
Two solutions of the 
homotopy Maurer-Cartan equation are identified if they are 
related by the action of $g_W$.

The   `homotopy   Maurer-Cartan  problem'   was   studied  in   recent
mathematical   work  of  M.~Kontsevich    \cite{Kontsevich_Felder, 
Kontsevich_Soibelman}  and
S.~Merkulov   \cite{Merkulov_defs,   Merkulov_infty}.   Upon  combining
their results with a simple  property of taking commutators (proved in
Appendix A), we show  that the two deformation problems (Maurer-Cartan
and homotopy-Maurer-Cartan)  give equivalent descriptions  of the same
moduli  space. This  explains the  relation between  the  string field
theory approach and the low energy point of view advocated in
 \cite{Douglas_quintic}. The fact that a homotopy version of the Maurer-Cartan 
equation arises  naturally in our  context suggests a  deeper relation
between      the       `derived      deformation      program'      of
 \cite{Kapranov_derived_defs}  and string field  theory. It  also sheds
light on the relation between D-brane superpotentials and the abstract
methods currently used in  the homological mirror symmetry literature.
In Section 4, we apply  our construction to the topological B/A models
which describe D-branes  wrapping a Calabi-Yau  manifold, respectively one
of  its   three-cycles,  thereby  obtaining  an   explicit  all  order
construction  of the  associated  D-brane superpotentials  (for the  A
model we only consider the  large radius limit, since the string field
theory at finite radius requires a more sophisticated analysis
 \cite{Fukaya, Fukaya2, boundary}). Section 5 presents our conclusions. 
Appendix A 
collects  some  facts  about  homotopy associative  and  homotopy  Lie
algebras and  their deformation theory  and proves a result  needed in
the  body  of  the  paper.  Appendix  B  contains  the  details  of  a
calculation relevant for understanding the effective symmetry algebra.

For the mathematically-oriented reader, I mention that many of the 
arguments used in this paper are adaptations 
of results known in the homological 
mirror symmetry literature. Unfortunately, they do not seem to have been 
integrated with each other and with the string field theory perspective, 
which is why their physical significance has remained somewhat obscure.

\section{Tree level potentials in open string field theory}

We start by presenting a method for computing tree level potentials in
(cubic) open string  field theory, and analyze the  result in terms of
$A_\infty$  algebras.    More  precisely,  we  show   that  a  certain
gauge-fixing  procedure leads  to a  collection of  tree  level string
products  which satisfy the  constraints of  an $A_\infty$  algebra as
well  as certain  cyclicity  properties. The  construction  we use  is
intimately  related  to the work of  \cite{Polishchuk}  and
 \cite{Kontsevich_Soibelman}, and we shall  borrow some of their results, 
with certain modifications.
While we shall  later 
apply this construction to  topological A/B strings
(see Section 4),  we chose to present it in an  abstract form in order
to  display the  complete similarity  between the  two theories.  
The procedure discussed below can also be applied to more general models, 
such as the graded string field theories of  \cite{com3,Diaconescu, sc,bv}.

\subsection{The abstract model}

Let us start with a cubic (topological) string field theory based on the action:
\be
\label{action}
S[\phi]=\frac{1}{2}\langle \phi, Q\phi\rangle+\frac{1}{3}\langle \phi, 
\phi\bullet \phi\rangle~~, 
\ee
where the string field $\phi$ is a degree one element of 
the boundary space ${\cal H}$, a $\Z$-graded differential associative algebra 
with respect to the (degree one) 
BRST operator $Q$, the boundary string product $\bullet$
and the worldsheet degree \footnote{For the topological 
A/B models, $|~.~|$ is the charge with respect to the anomalous $U(1)$ 
current on the worldsheet.} $|~.~|$. 
Since we shall deal with a single boundary sector (a single D-brane), we do 
not need to consider a category structure on ${\cal H}$ as in 
 \cite{com1, com3}. Our arguments can 
be generalized to that case, but in this note I wish to keep things simple.
Remember from  \cite{Witten_SFT, Thorn, com1} that the non-degenerate 
bilinear form 
$\langle .,.\rangle$ is invariant with respect to $Q$ and the boundary 
product:
\be
\label{metric_invar}
\langle Qu, v\rangle=-(-1)^{|u|}\langle u, Qv\rangle~~,~~
\langle u\bullet v, w\rangle=\langle u, v\bullet w\rangle~~.\nn
\ee 
It also has the graded symmetry property:
\be
\label{graded_symm}
\langle v,u\rangle =(-1)^{|u||v|}\langle v, u \rangle~~,
\ee
and obeys the selection rule:
\be
\langle u, v \rangle =0~~{\rm~unless~}~~|u|+|v|=3~~.
\ee
Due to this selection rule, the sign factor $(-1)^{|u||v|}$ in equation
(\ref{graded_symm}) can always be taken to be $+1$. 

\subsection{Gauge-fixing data}

We further assume that we are given a positive-definite Hermitian product 
$h$ on ${\cal H}$, which is antilinear with respect to its first variable
and couples only states of equal worldsheet degree:
\be
h(u,v)=0~~{\rm~unless~}~~|u|=|v|~~.
\ee
We let $Q^+$ be the Hermitian conjugate of $Q$ with respect to $h$:
\be
h(Qu, v)=h(u, Q^+v)~~.
\ee
Note that $Q^+$ is nilpotent and homogeneous of degree $-1$.

Let us define an antilinear operator $c$ on ${\cal H}$ through the relation:
\be
h(u,v)=\langle cu, v\rangle~~.
\ee
Since $\langle cu, v \rangle$ vanishes unless $|cu|+|v|=3$, while 
$h(u,v)$ vanishes unless $|v|-|u|=0$, we must have:
\be
|cu|=3-|u|
\ee
on homogeneous elements $u$.

Hermicity of $h$ is then equivalent with the property:
\be
\overline{\langle cu, v \rangle} =\langle cv, u \rangle~~, 
\ee
which is easily seen to imply:
\be
\label{c2}
\langle c^2u, v\rangle =\langle u, c^2v \rangle~~.
\ee
We shall {\em assume} that the metric $h$ is chosen such that 
$c^2=Id$ (this is possible in the topological A/B models, as 
we shall see in Section 4). With this hypothesis, it is easy to see that $c$ 
is an antilinear isometry with respect to $h$:
\be
h(cu,cv)=h(v,u)~~,
\ee
and that the operator $Q^+$ can be expressed as:
\be
\label{qd}
Q^+u=(-1)^{|u|}cQcu~~.
\ee
Indeed, one has:
\be
 (-1)^{|u|}h(cQcu, v)=(-1)^{|u|}\langle Qcu,v\rangle=
(-1)^{1+|u|+|cu|}\langle cu, Qv\rangle=\langle cu, Qv\rangle
=h(u, Qv)~~.
\ee
Using (\ref{qd}), one can check that the defining relation 
for $Q^+$ (namely $h(Qu, v)=h(u, Q^+v)$) implies:
\be
\label{qdag}
\langle Q^+u, v\rangle =(-1)^{|u|}\langle u, Q^+v\rangle~~.
\ee
This property will be essential in Subsection 2.5.

\subsection{The propagator}

The string field action (\ref{action}) has the gauge symmetry:
\be
\phi\rightarrow \phi-Q\alpha-[\phi,\alpha]~~,
\ee
with $\alpha$ a degree zero element of ${\cal H}$. 
We are interested in partially fixing this symmetry through the gauge 
condition
\footnote{For the topological A/B models discussed in Section 4, this 
is essentially the Siegel gauge (as follows from the 
fact that in such models one can identify $b_0$ with $Q^+$ for an 
appropriate choice of the metric $h$), hence 
the associated string field correlators admit a direct interpretation
as string scattering amplitudes.}:
\be
\label{Gauge}
Q^+\phi=0~~.
\ee
A thorough analysis of gauge fixing would generally require 
the full machinery of the BV formalism, but luckily we 
will not need this here. In fact, we shall only be interested 
in tree level scattering amplitudes for the topological A/B models, 
for which it suffices to understand the relevant propagator. 
That the BV analysis does not modify the discussion in this 
case follows, for example, from the work of  \cite{AS}.

For this purpose it is convenient to consider 
the `Hodge theory' of $Q$. Let us define\footnote{In this note, 
$[ . , . ]$ always stands for the graded commutator.}:
\be
H=[Q,Q^+]=QQ^++Q^+Q~~,
\ee
and let $K$ denote the kernel of $H$. As usual in Hodge theory, one has:
\bea
\label{hodge}
K=Ker Q\cap Ker Q^+~~&,&~~{\cal H}=K\oplus ImQ\oplus ImQ^+~~,\\
Ker Q=K\oplus Im Q~~~~~&,&~~~~~ Ker Q^+=K\oplus Im Q^+,
\eea
where the direct sums involved are orthogonal with respect to $h$. 
The operator $H$ is Hermitian and commutes with $Q$ and $Q^+$, 
and thus its restriction to $K^\perp=Im Q \oplus ImQ^+$ 
gives an automorphism of this space. 
We shall denote the inverse of $H|_{K^\perp}$ by $\frac{1}{H}$. 
It is easy to check that the operators $\pi_Q=Q\frac{1}{H}Q^+$ 
and $\pi_{Q^+}=Q^+\frac{1}{H}Q$ are orthogonal projectors on 
$Im Q$ and $Im Q^+$. It follows that the operator:
\bea
P&=& 1-(Q^+\frac{1}{H}Q+Q\frac{1}{H}Q^+)~~\\
\eea 
is the orthogonal projector on $K$. 

To find the relevant propagator, one must identify a maximal subspace 
of ${\cal H}^1$ on which the quadratic form 
$S_{kin}(\phi)=\langle \phi, Q\phi \rangle$ is non-degenerate. 
Since the BPZ form $\langle .,. \rangle$ is non-degenerate, 
the kernel of the bilinear symmetric form $\langle \phi, Q\psi\rangle=
\langle \psi, Q\phi\rangle=\langle Q\phi, \psi \rangle= h(cQ\phi, \psi)=
- h(Q^+(c\phi), \psi)$ on ${\cal H}^1$
(the polar form of $S_{kin}$) is 
$(ImQ^+)^\perp\cap {\cal H}^1=KerQ\cap {\cal H}^1$.
Hence a maximal subspace with the desired property is 
$(KerQ)^\perp \cap {\cal H}^1=ImQ^+\cap {\cal H}^1$. The restriction of 
$Q$ to $Im Q^+$ gives an isomorphism from $Im Q^+$ to $Im Q$, whose inverse 
we denote by $\frac{1}{Q}$. It follows that for 
$\phi\in Im Q^+\cap {\cal H}^1$ we can write:
\be
\langle \phi, Q\phi\rangle=\langle \psi, \frac{1}{Q}\psi\rangle~~,
\ee
where $\psi=Q\phi$. Hence the pull-back of $S_{kin}$ through the map 
$\frac{1}{Q}:Im Q\cap {\cal H}^2\stackrel{\approx}{\rightarrow} 
Im Q^+\cap {\cal H}^1$ can be identified 
with its `inverse'.

Let us consider a `Green operator' $G$ for $H$, which satisfies:
\bea
HG=1-P=\pi_Q+\pi_{Q^+}~~.
\eea
We shall choose the solution 
$G=\frac{1}{H}(1-P)=\frac{1}{H}(\pi_{Q}+\pi_{Q^+})$. 
We next 
define $U=Q^+G=Q^+\frac{1}{H}(\pi_{Q}+\pi_{Q^+})=Q^+\frac{1}{H}\pi_{Q}=
\frac{1}{H}Q^+$.
Then the projectors $\pi_Q, \pi_{Q^+}$ can be written as:
\bea
\label{foo1}
\pi_Q=QU~~,~~\pi_{Q^+}=UQ~~.
\eea
Hence we can write:
\be
U=\frac{1}{Q}\pi_Q.
\ee
It is clear that $U$ plays the role of propagator 
for the $Q$-exact modes. Following the terminology of 
 \cite{Zwiebach_closed}, states belonging to the subspace 
$ImQ$ will be called {\em spurious}, while states belonging to 
$K=KerQ\cap Ker Q^+$ will be called {\em physical}\footnote{Strictly speaking, 
physical states are {\em degree one} elements of $K$, but we shall sometimes 
use the words `physical states' to mean states belonging to 
$K$. We hope this does not lead to confusion.}. The elements 
of $KerQ^\perp=ImQ^+$ are the {\em unphysical} states. 
Relations 
(\ref{hodge}) show that the off-shell state space ${\cal H}$ 
decomposes into physical, spurious and unphysical components.
It is clear that $U$ propagates spurious states into unphysical states
and projects out everything else. 

\subsection{Tree level amplitudes and the potential} 

It may seem strange that we are interested in a propagator which describes 
the dynamics of non-physical states. The reason why such an object is 
relevant is that {\em the basic string product $\bullet$ does not 
map physical states into physical states}. Indeed, the string field theory 
axioms assure that $\bullet$ maps $Ker Q\times Ker Q$ into $Ker Q$ 
(since the BRST operator acts as a derivation of the product), but it is 
{\em not} 
true, in general, that $\bullet$ maps $Ker Q^+\times Ker Q^+$ into $Ker Q^+$
(in particular, $Q^+$ does not act as a derivation of the string product, 
even though it has property (\ref{qdag}) with respect to the bilinear form).
If one considers a two-string joining process 
$(u_1, u_2)\rightarrow v=u_1\bullet u_2$, then the state $v$ will generally not satisfy 
the gauge-fixing condition $Q^+v=0$, even if both $u_1$ and $u_2$ belong to 
the space of physical states $K^1$. Since $Qv=0$, we have $v\in Ker Q=K\oplus
Im Q$, so the precise way in which the gauge condition is violated is that 
$v$ may acquire a component $v_Q\in Im Q$ along the subspace of spurious states.
This component then propagates into the unphysical state 
$Uv_Q=Uv\in Im Q^+$. 
If the composite string now  
interacts with an open string in the state $u_3\in K^1$, the result is 
$(Uv)\bullet u_3=(U(u_1\bullet u_2))\bullet u_3$, which can be measured 
by projecting onto $K$ etc. It follows that string amplitudes 
$\langle \langle u_1\dots u_n\rangle\rangle$, where $u_1\dots u_n$ are (degree one) physical 
states, are built according to the Feynman rules of the cubic theory 
(\ref{action}) upon using the propagator $U$. To be precise, we define 
$\langle \langle u_1\dots u_n\rangle\rangle^{(n)}$ to be {\em amputated} 
amplitudes, so there are no insertions of propagators on the 
external legs. Moreover, we shall only be interested in tree level 
correlators, 
which we denote by $\langle \langle u_1\dots u_n\rangle\rangle_{tree}^{(n)}$.

We next define a tree-level potential by summing over all (signed) 
amputated (not necessarily connected) 
tree-level scattering amplitudes with at least three legs:
\be
\label{superpot}
W[\phi]=\sum_{n\geq 3}{\frac{1}{n}(-1)^{n(n-1)/2}
\langle \langle \phi,\dots ,\phi\rangle \rangle^{(n)}_{tree}}~~.
\ee
The string field $\phi$ in this expression belongs to 
the space $K^1=K\cap {\cal H}^1$. 

It is convenient formulate this in an algebraic manner. 
Let us define string products $r_n:K^{\otimes n}\rightarrow K$
$(n\ge 2)$ 
by following the tree level Feynman diagrams of our theory, but applied 
to propagation of arbitrary states $u\in K$ (i.e. we formally allow 
$u$ to have degree different from one). Upon following the combinatorics of 
tree level diagrams, one can easily check that $r_n$ can be 
described as follows (figure 1):

\

1. We first define products $\lambda_n:{\cal H}^n\rightarrow {\cal H}$ 
through $\lambda_2=\bullet$ and the recursion relation:
{\footnotesize \bea
\label{rec}
\lambda_n(u_1,\dots ,u_n)&=&(-1)^{n-1}
(U\lambda_{n-1}(u_1,\dots ,u_{n-1}))u_n-(-1)^{n|u_1|}
u_1(U\lambda_{n-1}(u_2,\dots ,u_n))-\nn\\
&&\sum_{\tiny \begin{array}{c}k+l=n\\ k,l\ge 2\end{array}}
(-1)^{k+(l-1)(|u_1|+\dots +|u_k|)}
(U\lambda_k(u_1,\dots ,u_k))(U\lambda_l(u_{k+1},\dots ,u_n))~~,
\eea}
for $u_1\dots u_n$ in ${\cal H}$. 

\

2. The products $r_n$ are then given by:
\be
\label{rs}
r_n(u_1,\dots ,u_n)=P\lambda_n(u_1,\dots ,u_n)~~,
\ee
for $u_1, \dots  , u_n\in K$.

\

The recursion relation (\ref{rec}) describes the decomposition of 
an order $n$ tree level product 
into lower order products, as explained in 
figure 1. This  
encodes the combinatorics of tree level Feynman diagrams.
With our conventions for the grading, the product $r_n$ has degree $2-n$:
\be
|r_n(u_1\dots u_n)|=|u_1|+\dots |u_n|+2-n~~.
\ee

\hskip 1.0 in
\begin{center} 
\scalebox{0.5}{\begin{picture}(0,0)%
\includegraphics{feynman.pstex}%
\end{picture}%
\setlength{\unitlength}{4144sp}%
\begingroup\makeatletter\ifx\SetFigFont\undefined%
\gdef\SetFigFont#1#2#3#4#5{%
  \reset@font\fontsize{#1}{#2pt}%
  \fontfamily{#3}\fontseries{#4}\fontshape{#5}%
  \selectfont}%
\fi\endgroup%
\begin{picture}(9192,7167)(46,-6631)
\put(2611,-376){\makebox(0,0)[lb]{\smash{\SetFigFont{17}{20.4}{\familydefault}{\mddefault}{\updefault}
$=$
}}}
\put(3421,-2131){\makebox(0,0)[lb]{\smash{\SetFigFont{17}{20.4}{\familydefault}{\mddefault}{\updefault}
$u_1$
}}}
\put(5356,-1456){\makebox(0,0)[lb]{\smash{\SetFigFont{17}{20.4}{\familydefault}{\mddefault}{\updefault}
$u_3$
}}}
\put(4636,-2131){\makebox(0,0)[lb]{\smash{\SetFigFont{17}{20.4}{\familydefault}{\mddefault}{\updefault}
$u_2$
}}}
\put(5041,-61){\makebox(0,0)[lb]{\smash{\SetFigFont{17}{20.4}{\familydefault}{\mddefault}{\updefault}
$P$
}}}
\put(4096,-781){\makebox(0,0)[lb]{\smash{\SetFigFont{17}{20.4}{\familydefault}{\mddefault}{\updefault}
$U$
}}}
\put(7156,-1321){\makebox(0,0)[lb]{\smash{\SetFigFont{17}{20.4}{\familydefault}{\mddefault}{\updefault}
$u_1$
}}}
\put(7786,-2041){\makebox(0,0)[lb]{\smash{\SetFigFont{17}{20.4}{\familydefault}{\mddefault}{\updefault}
$u_2$
}}}
\put(9136,-2131){\makebox(0,0)[lb]{\smash{\SetFigFont{17}{20.4}{\familydefault}{\mddefault}{\updefault}
$u_3$
}}}
\put(8551,-691){\makebox(0,0)[lb]{\smash{\SetFigFont{17}{20.4}{\familydefault}{\mddefault}{\updefault}
$U$
}}}
\put(8326,-61){\makebox(0,0)[lb]{\smash{\SetFigFont{17}{20.4}{\familydefault}{\mddefault}{\updefault}
$P$
}}}
\put(5671,-376){\makebox(0,0)[lb]{\smash{\SetFigFont{17}{20.4}{\familydefault}{\mddefault}{\updefault}
$-(-1)^{|u_1|}$
}}}
\put(181,-6139){\makebox(0,0)[lb]{\smash{\SetFigFont{14}{16.8}{\familydefault}{\mddefault}{\updefault}
$u_1$
}}}
\put(978,-6382){\makebox(0,0)[lb]{\smash{\SetFigFont{14}{16.8}{\familydefault}{\mddefault}{\updefault}
$u_2$
}}}
\put(8506,-5191){\makebox(0,0)[lb]{\smash{\SetFigFont{14}{16.8}{\familydefault}{\mddefault}{\updefault}
$U$
}}}
\put(7561,-5191){\makebox(0,0)[lb]{\smash{\SetFigFont{14}{16.8}{\familydefault}{\mddefault}{\updefault}
$U$
}}}
\put(8416,-4786){\makebox(0,0)[lb]{\smash{\SetFigFont{14}{16.8}{\familydefault}{\mddefault}{\updefault}
$P$
}}}
\put(7381,-6586){\makebox(0,0)[lb]{\smash{\SetFigFont{14}{16.8}{\familydefault}{\mddefault}{\updefault}
$u_2$
}}}
\put(6751,-5776){\makebox(0,0)[lb]{\smash{\SetFigFont{14}{16.8}{\familydefault}{\mddefault}{\updefault}
$u_1$
}}}
\put(7606,-6046){\makebox(0,0)[lb]{\smash{\SetFigFont{14}{16.8}{\familydefault}{\mddefault}{\updefault}
$u_{k}$
}}}
\put(8191,-6271){\makebox(0,0)[lb]{\smash{\SetFigFont{14}{16.8}{\familydefault}{\mddefault}{\updefault}
$u_{k+1}$
}}}
\put(8641,-6631){\makebox(0,0)[lb]{\smash{\SetFigFont{14}{16.8}{\familydefault}{\mddefault}{\updefault}
$u_{k+2}$
}}}
\put(9226,-6226){\makebox(0,0)[lb]{\smash{\SetFigFont{14}{16.8}{\familydefault}{\mddefault}{\updefault}
$u_n$
}}}
\put(1801,-6361){\makebox(0,0)[lb]{\smash{\SetFigFont{14}{16.8}{\familydefault}{\mddefault}{\updefault}
$u_n$
}}}
\put(3196,-5551){\makebox(0,0)[lb]{\smash{\SetFigFont{41}{49.2}{\familydefault}{\mddefault}{\updefault}
${\huge =\sum_k{\pm}}$
}}}
\put( 46,-1501){\makebox(0,0)[lb]{\smash{\SetFigFont{17}{20.4}{\familydefault}{\mddefault}{\updefault}
$u_1$
}}}
\put(1081,-1816){\makebox(0,0)[lb]{\smash{\SetFigFont{17}{20.4}{\familydefault}{\mddefault}{\updefault}
$u_2$
}}}
\put(2116,-1456){\makebox(0,0)[lb]{\smash{\SetFigFont{17}{20.4}{\familydefault}{\mddefault}{\updefault}
$u_3$
}}}
\end{picture}
}
\end{center}
\begin{center} 
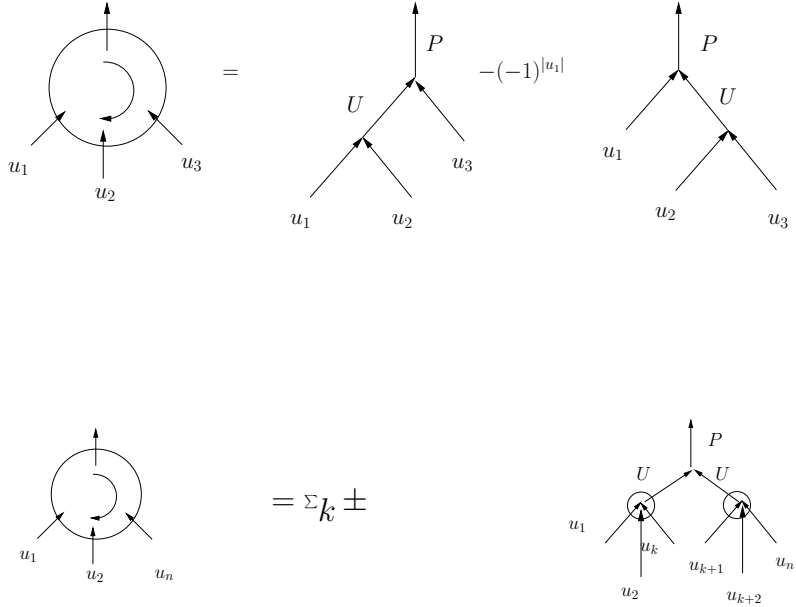
Figure  1. {\footnotesize Expressing 
disk string correlators of physical states in terms of Feynman rules.
The upper figure shows the case of the product $r_3=P\left[U(u_1\bullet u_2)
\bullet u_3-(-1)^{|u_1|}u_1\bullet U(u_2\bullet u_3)\right]$. The lower figure 
shows the general decomposition of $\lambda_n$ with respect to 
the products $\lambda_k$ ($k< n$).}
\end{center}

Products of the type (\ref{rs}) were considered in
 \cite{Polishchuk},  \cite{Merkulov} and 
 \cite{Kontsevich_Soibelman}. In those papers, it is shown that they define 
an $A_\infty$ algebra structure on $K$, i.e. they satisfy the following 
constraints:
\be
\label{A_infty}
\sum_{\tiny \begin{array}{c}k+l=n+1\\j=0\dots k-1\end{array}}
{(-1)^s r_k(u_1\dots u_j,r_l(u_{j+1}\dots u_{j+l}),u_{j+l+1}\dots u_n)}=0~~,
\ee
for all $n\geq 3$, where $s=l(|u_1|+\dots |u_j|)+j(l-1)+(k-1)l$.
Note that our algebra has $r_1=0$. Some basic facts about $A_\infty$ 
algebras are collected in Appendix A.

With these preparations, we define (extended) tree level amplitudes by:
\be
\label{amplitudes}
\langle \langle u_1\dots u_n\rangle \rangle^{(n)}_{tree}=
\langle u_1, r_{n-1}(u_2\dots u_n)\rangle~~,
\ee
where $u_1\dots u_n$ belong to $K$.
Expression
(\ref{amplitudes}) makes mathematical sense for elements of $K$ of arbitrary 
degree.
With our definition of $r_n$, the quantities (\ref{amplitudes}) 
coincide with the amputated tree level amplitudes when 
$u_1\dots u_n$ are degree one elements of $K$. 
Hence we can write our potential as follows:
\be
\label{W}
W(\phi)=\sum_{n\geq 2}{\frac{1}{n+1}(-1)^{n(n+1)/2}
\langle\phi, r_n(\phi^{\otimes n})
\rangle}~~.
\ee

\paragraph{Observation}

In a topological string theory, the potential $W$ can be identified 
with a certain representation of the open string analogue of the 
`free energy' of  \cite{DVV}, which was studied from a deformation-theoretic 
perspective in  \cite{Hofman1}. To understand this relation, let us 
pick a basis $\phi_1\dots \phi_h$ of $K^1$ and look for the value of 
our potential at the point $\phi=\sum_{j=1}^h{t_j\phi_j}$, where 
$t=(t_1\dots t_h)$ is a collection of parameters:
\be
W(t)=\sum_{s_1\dots s_h\geq 0}{t_1^{s_1}\dots t_h^{s_h}W_{s_1\dots s_h}}(\phi_1\dots \phi_h)~~.
\ee
Here $W_{s_1\dots s_h}(\phi_1\dots \phi_h)$ involves a sum of string amplitudes 
with $s_j$ insertions of $\phi_j$ in all possible orderings on the 
disk's boundary. Upon fixing the position of three distinct insertion  points, 
those coefficients $W_{s_1\dots s_h}$ for which $s_1+\dots +s_h\geq 3$ can be 
formally written in terms of integrated descendants. It is then easy to see 
that this corresponds to the boundary potential considered in  \cite{Hofman1}.
Our description differs from that of  \cite{Hofman1} in that it uses a 
gauge-fixing prescription in order to give a concrete formula for 
string amplitudes. As we shall see below, this allows one to give a 
more detailed analysis of boundary deformations, 
which goes beyond the infinitesimal
(first order) approach. In general, it seems that progress in the study of 
deformations requires the full force of string field theory.

\subsection{Cyclicity}

It is possible to show 
that our tree level correlators satisfy the 
following cyclicity property:
\be
\label{cycl}
\langle \langle u_1\dots u_n\rangle \rangle^{(n)}_{tree}=
(-1)^{(n-1)(|u_1|+|u_2|+1)+|u_1|(|u_2|+\dots +|u_n|)}
\langle \langle u_2\dots u_n, u_1\rangle\rangle^{(n)}_{tree}~~,
\ee
i.e.:
\be
\label{cc1}
\langle u_1, r_n(u_2\dots u_{n+1})\rangle=(-1)^{n(|u_1|+|u_2|+1)+
|u_1|(|u_2|+\dots +|u_{n+1}|)}\langle u_2, r_n(u_3\dots u_{n+1},u_1)\rangle~~.
\ee
For this, note that (\ref{qdag}) implies that 
the operator $U$ has a similar property:
\be
\langle Uu, v\rangle =(-1)^{|u|}\langle u, Uv\rangle~~.
\ee
The rest of the 
argument is then formally identical\footnote{The abstract form of the 
argument of \cite{Polishchuk} can be most easily recovered upon 
defining the `trace' $Tr(u):=\langle u,1\rangle =\langle 1, u\rangle$
on ${\cal H}$, where $1$ is the unit of the boundary algebra 
$({\cal H},\bullet)$. Invariance of the bilinear form
with respect to the boundary product 
implies $\langle u,v\rangle=Tr(u\bullet v)$, which allows one to 
apply the cyclicity argument of \cite{Polishchuk} to our more 
general situation.} with that given in  \cite{Polishchuk}, 
and will not be repeated here. 

We note that the selection rule for the bilinear form 
$\langle .,.\rangle$ allows one to 
simplify (\ref{cc1}) to:
\be
\label{cc2}
\langle u_1, r_n(u_2\dots u_{n+1})\rangle=(-1)^{n(|u_2|+1)}
\langle u_2, r_n(u_3\dots u_{n+1},u_1)\rangle~~.
\ee

\section{Two descriptions of the boundary moduli space}
In this section we give two descriptions of the moduli space
of vacua. The first is the standard 
construction in terms of solutions of the string field equations of 
motion, while the second results by considering extrema of the potential 
$W$. We shall show that the two descriptions are locally equivalent 
by formulating them in terms of Lie/homotopy Lie algebras 
and using mathematical results of M.~Kontsevich and S.~A.~Merkulov.
Some ideas of this section are already implicit in  \cite{Witten_CS}.

\subsection{The string field theory description}

The space ${\cal M}$ of vacua of a cubic string field theory can 
be described as the moduli space of degree one 
solutions to the Maurer-Cartan equations (=string field equations of motion)
\be
\label{mc}
Q\phi+\frac{1}{2}[\phi,\phi]=0 ~~,
\ee
taken modulo the action of the gauge group ${\cal G}$ 
generated by transformations of the form:
\be
\label{gauge_infin}
\phi\rightarrow \phi-Q\alpha-[\phi,\alpha]~~,
\ee
where the infinitesimal generator $\alpha$ is a degree zero element of 
${\cal H}$.
In these equations, $[.,.]$ stands for the graded commutator 
in the graded associative algebra ${\cal H}$:
\be
\label{comm}
[u,v]:=u\bullet v -(-1)^{|u||v|}v\bullet u~~.
\ee
The gauge group ${\cal G}$ can be described globally as follows. 
When endowed with the commutator (\ref{comm}), the space 
${\cal H}$ 
becomes a differential graded Lie algebra ${\bf g}$; 
the relation between this  
and the graded associative algebra $({\cal H}, \bullet)$ is entirely similar 
to that between a usual (ungraded) associative algebra and the corresponding 
Lie algebra. It is easy to see that the subspace ${\cal H}^0$ of 
degree zero elements forms an (ungraded) Lie sub-algebra 
$g:={\bf g}^0$ of 
${\bf g}=({\cal H}, [.,.])$; this coincides with the commutator algebra 
of the (ungraded) associative subalgebra $({\cal H}^0,\bullet)$.
The gauge group ${\cal G}$ is formally the Lie group 
obtained by exponentiating this Lie algebra. It consists of elements 
$\lambda=exp(\alpha)$, where $exp$ denotes the exponential map. This 
description is only formal because, even in the simplest case of topological 
string theories, the Lie algebra ${\bf g}$ is in fact infinite-dimensional, 
and thus the exponential has to be carefully defined on a case by case basis. 
The group ${\cal G}$ acts on the space ${\cal H}$ though 
the obvious extension of its adjoint representation:
\be
e^\alpha\bullet u=e^{ad_\alpha}u~~,
\ee
where $ad$ is the adjoint action of the Lie algebra $g$:
\be
ad_\alpha(u)=[\alpha, u]~~.
\ee
Under the action of $e^\alpha$, the 
string field $\phi$ is taken to transform as a `connection':
\be
\label{gauge}
\phi\rightarrow \phi^\alpha=e^{ad_\alpha}\phi-
\frac{e^{ad_\alpha}-1}{ad_\alpha}Q\alpha~~,
\ee
where the last term is defined through its series
expansion:
\be
\frac{e^{ad_\alpha}-1}{ad_\alpha}=\sum_{n\geq 1}{\frac{1}{n!}(ad_\alpha)^{n-1}}~~.
\ee
Upon expanding (\ref{gauge}) to first order in $\alpha$, one recovers the 
infinitesimal form (\ref{gauge_infin}).

\subsection{Description through extrema of the potential}

The description of the moduli space discussed above   
displays the complete analogy between cubic string field theory and 
Chern-Simons field theory. It is possible to give an entirely 
different construction, which is based on the potential 
(\ref{W}). Indeed, one can ask for the moduli space 
${\cal M}_W$ of string field configurations $\phi\in K^1$ 
which extremize this potential:
\be
\label{mc_W}
\frac{\partial W}{\partial \phi}(\phi)=0
\Leftrightarrow \sum_{n\geq 2}{(-1)^{n(n+1)/2}r_n(\phi^{\otimes n})}=0~~.
\ee
To arrive at this equation, we noticed that the 
cyclicity property (\ref{cycl}) implies:
\be
\langle \langle u_1\dots u_n\rangle \rangle^{(n)}_{tree}=
\langle \langle u_2\dots u_n, u_1
\rangle \rangle^{(n)}_{tree}~{\rm~for~}~u_1\dots u_n\in {\cal H}^1~~.
\ee

\subsubsection{Formulation of the extremum condition in terms of an $L_\infty$ 
algebra}

In order to understand the relevant algebra of symmetries, 
it is convenient to rewrite equation 
(\ref{mc_W}) 
in terms of the `commutator algebra' of the $A_\infty$ algebra 
defined by the products $r_n$. Just as any (differential) graded 
associative algebra defines a (differential) graded Lie algebra, any 
$A_\infty$ algebra  
has an associated $L_\infty$ (or {\em strong homotopy Lie}) 
algebra. To describe this construction, 
we first recall the definition of 
these mathematical structures (the reader can find more details in 
Appendix A).

\paragraph{$L_\infty$ algebras}

$L_\infty$ algebras $(L, \{m_n\}_{n\geq 1})$ are natural 
generalizations of Lie algebras, being defined by a countable family of 
$n$-tuple products $m_n$ subject to the constraints:
\be
\label{L_inf}
\sum_{k+l=n+1}\sum_{\sigma\in Sh(k,n)}{(-1)^{k(l-1)}\chi(\sigma; u_1\dots u_n)
m_l(m_k(u_{\sigma(1)}\dots u_{\sigma(k)}),u_{\sigma(k+1)}\dots u_{\sigma(n)})}=0~~,
\ee
and to the graded antisymmetry condition:
\be
m_n(u_{\sigma(1)}\dots u_{\sigma(n)})=\chi(\sigma; u_1\dots u_n)m_n(u_1\dots u_n)~~,
\ee
for any permutation $\sigma$ of the set $\{1\dots n\}$. 
In these relations, the symbol 
$\chi(\sigma, u_1\dots u_n)=\pm 1$ 
is defined through:
\be
\chi(\sigma;u_1\dots u_n):=\epsilon(\sigma)\epsilon(\sigma; u_1\dots u_n)~~,
\ee
where $\epsilon(\sigma)$ is the signature of the permutation $\sigma$ and 
$\epsilon(\sigma; u_1\dots u_n)$ is the so-called {\em Koszul sign}, i.e.
the sign obtained when unshuffling the 
element $u_{\sigma(1)}\odot \dots \odot u_{\sigma(n)}$ to the form 
$u_1\odot \dots \odot u_n$ in the free graded commutative algebra 
$\odot^*L=\oplus_{k\geq 0}{\odot^kL}$:
\be
u_{\sigma(1)}\odot \dots \odot u_{\sigma(n)}=\epsilon(\sigma; u_1\dots u_n)
u_1\odot \dots \odot u_n~~.
\ee
We remind the reader that $\odot^* L $ is built upon dividing the free 
associative algebra 
$\otimes^*L=\oplus_{k\geq 0}{\otimes^kL}$ through the homogeneous 
ideal generated by elements of the form $u\otimes v-(-1)^{|u||v|}v\otimes u$.

The sum in (\ref{L_inf}) is over so-called {\em $(k,n)$-shuffles}, i.e. 
permutations $\sigma$ on $n$ elements which satisfy:
\be
\sigma(1)<\sigma(2)<\dots <\sigma(k)~~,~~
\sigma(k+1)<\sigma(k+2)<\dots <\sigma(n)~~.
\ee

An $L_\infty$ algebra such that $m_n=0$ for all $n\geq 3$ is simply 
a differential graded Lie algebra, with the differential $Q=m_1$ and 
the graded Lie bracket $[.,.]=m_2$.

\paragraph{The commutator algebra of an $A_\infty$ algebra}

Given an $A_\infty$ algebra $(A, \{r_n\}_{n\geq 1})$, its {\em commutator
algebra}  \cite{Lada_Markl} is  the the $L_\infty$ algebra
defined on the same underlying space $L=A$ by the products:
\be
\label{commutator}
m_n(u_1\dots u_n)=\sum_{\sigma\in S_n}{\chi(\sigma, u_1\dots u_n)r_n(u_{\sigma(1)}\dots 
u_{\sigma(n)})}~~,
\ee
where $S_n$ is the permutation group on $n$ elements. It is easy to check 
by computation 
that the defining constraints of an $L_\infty$ algebra are satisfied. 
A more synthetic description of this construction (in terms of so-called 
bar duals) can be found in  \cite{Lada_Markl} and is summarized in Appendix A.

\paragraph{The homotopy Maurer-Cartan problem} 

Let us return to equations (\ref{mc_W}). 
Performing the commutator construction for our $A_\infty$ algebra
$(K, r_n)$, we 
obtain an $L_\infty$ algebra $(K, m_n)$ whose first product $m_1$ vanishes. 
If we apply (\ref{commutator}) to $u_1=\dots =u_n=\phi$, we obtain:
\be
m_n(\phi\dots \phi)=n!r_n(\phi\dots \phi)~~,
\ee
where we used the fact that $\epsilon(\sigma, \phi\dots \phi)=\epsilon(\sigma)$
(and thus $\chi(\sigma, \phi\dots \phi)=+1$), which follows from $|\phi|=1$.
Hence one can rewrite the extremum conditions 
(\ref{mc_W}) as:
\be
\label{mc_L}
\sum_{n\geq 2}{\frac{(-1)^{n(n+1)/2}}{n!}m_n(\phi^{\otimes n})}=0~~.
\ee
This is the standard form of the so-called 
`homotopy Maurer-Cartan equation' in an 
$L_\infty$ algebra  \cite{Kontsevich_Felder, Merkulov_infty}. 
It is possible to check (see Appendix B) that equations (\ref{mc_L}) are 
invariant with respect to infinitesimal transformations of the form:
\be
\label{gauge_W}
\phi\rightarrow \phi'=\phi+\delta_\alpha\phi~, {\rm~with~}~
\delta_\alpha\phi=-\sum_{n\geq 2}{\frac{(-1)^{n(n-1)/2}}{(n-1)!}
m_n(\alpha\otimes \phi^{\otimes n-1})}~~,\nn
\ee
where $\alpha$ is a degree zero element of $K$\footnote{These transformations 
can be formulated abstractly in terms of so-called `pointed Q-manifolds'
 \cite{Kontsevich_Felder}. The explicit form given here is justified by the 
calculations of Appendix B. A similar explicit form is 
written down in  \cite{Merkulov_infty}, though 
the choice of signs in that paper
seems to differ from ours.}. The moduli space ${\cal M}_W$ is 
then defined by modding out the space of solutions to (\ref{mc_W}) 
or (\ref{mc_L}) through the action of the symmetry algebra 
$g_W$ generated by (\ref{gauge_W}).

\paragraph{Observations} 
(1) The basic difference between the algebras $g={\bf g}^0$ 
and $g_W$ is that the action of the former involves the BRST 
differential (equation (\ref{gauge_infin})), 
while the action of the latter does not. 
Passage from the string field theory to the 
tree level effective description `rigidifies'  
$g$ to $g_W$.

(2) The algebra of transformations (\ref{gauge_W}) is generally open, i.e. 
it only closes on the critical set of $W$ (this seems to happen for the 
graded string field theories of \cite{com3, Diaconescu, sc, bv}). 
For $\phi$ satisfying the 
critical point equations (\ref{mc_W}), the commutator
$\delta_\alpha\delta_\beta\phi-\delta_\beta\delta_\alpha\phi$ does not
generally coincide  
with $\delta_{[\alpha,\beta]}$, but with an infinitesimal transformation 
$\delta_\gamma\phi$, where $\gamma=\gamma(\alpha,\beta,\phi)$ 
is given by a sum over products of the form 
$m_n(\alpha\otimes \beta\otimes \phi^{n-2})$. This situation is 
familiar  in the context of the BV-BRST formalism. The structure of 
$g_W$ is 
much simpler for the ungraded string field theories 
discussed in Section 4. In this case, one can show that the gauge algebra 
closes away from the critical set of $W$ 
to a standard Lie algebra.  This follows from the 
argument given below. 

\subsubsection{A particular case}

Let us assume for a moment that:
\be
\label{der_dagger}
Q^+(\alpha\bullet u)=\alpha\bullet Q^+u {\rm~for~}\alpha\in K^0 {\rm~and~}u\in 
{\cal H}
\ee 
(this holds for the topological A/B models, 
which will be discussed 
below). With this assumption, one can show 
that the infinitesimal 
gauge transformations (\ref{gauge_W}) reduce to:
\be
\label{moo}
\phi\rightarrow \phi +[\alpha, \phi]~~.
\ee
To prove this, note that (\ref{der_dagger}) implies 
$U(\alpha\bullet u)=0$ for any $u$ which belongs to $KerQ^+$.

If we consider the diagrammatic expansion of the products $r_n$, 
this implies that a connected contribution 
to $r_n(\alpha,\phi^{\otimes n-1})$ (with $\alpha\in K^0$ 
and $\phi\in K^1$) vanishes unless both of the following conditions are 
satisfied:

\

(1) the $\alpha$-insertion belongs to the highest 
level of the associated tree (i.e. belongs to the same node 
as a $\phi$-insertion). This follows from the fact that any expression of 
the form $U(\alpha\bullet Uv)$ vanishes, since $Uv\in ImQ^+\subset Ker Q^+$. 
Hence branches of the type displayed in Figure 2(a) are forbidden.

\

(2) its insertion node is the root of the tree, i.e. the node where the 
projector $P$ is inserted. This follows from the fact that 
$U(\alpha\bullet \phi)=0$, since $\phi\in K^1\subset Ker Q^+$. 
Hence branches of the type displayed in Figure 2(b) are forbidden.

\

\hskip 1.0 in
\begin{center} 
\scalebox{0.5}{\begin{picture}(0,0)%
\includegraphics{branches.pstex}%
\end{picture}%
\setlength{\unitlength}{4144sp}%
\begingroup\makeatletter\ifx\SetFigFont\undefined%
\gdef\SetFigFont#1#2#3#4#5{%
  \reset@font\fontsize{#1}{#2pt}%
  \fontfamily{#3}\fontseries{#4}\fontshape{#5}%
  \selectfont}%
\fi\endgroup%
\begin{picture}(5190,3037)(811,-3976)
\put(1306,-3976){\makebox(0,0)[lb]{\smash{\SetFigFont{17}{20.4}{\familydefault}{\mddefault}{\updefault}
$(a)$
}}}
\put(5176,-3841){\makebox(0,0)[lb]{\smash{\SetFigFont{17}{20.4}{\familydefault}{\mddefault}{\updefault}
$(b)$
}}}
\put(4276,-3076){\makebox(0,0)[lb]{\smash{\SetFigFont{17}{20.4}{\familydefault}{\mddefault}{\updefault}
$\alpha$
}}}
\put(5851,-3076){\makebox(0,0)[lb]{\smash{\SetFigFont{17}{20.4}{\familydefault}{\mddefault}{\updefault}
$\phi$
}}}
\put(5581,-1501){\makebox(0,0)[lb]{\smash{\SetFigFont{17}{20.4}{\familydefault}{\mddefault}{\updefault}
$U$
}}}
\put(811,-3301){\makebox(0,0)[lb]{\smash{\SetFigFont{17}{20.4}{\familydefault}{\mddefault}{\updefault}
$\alpha$
}}}
\put(2026,-2131){\makebox(0,0)[lb]{\smash{\SetFigFont{17}{20.4}{\familydefault}{\mddefault}{\updefault}
$U$
}}}
\put(1981,-1591){\makebox(0,0)[lb]{\smash{\SetFigFont{17}{20.4}{\familydefault}{\mddefault}{\updefault}
$U$
}}}
\end{picture}
}
\end{center}
\begin{center} 
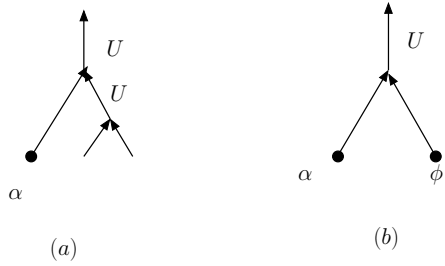
Figure  2. {\footnotesize Branches which lead to vanishing of a tree-level 
contribution to $r_n(\alpha,\phi^{\otimes n-1})$.  
The two edges on the right of figure (a) may be internal or external.}

\end{center}

It is clear that it is impossible to satisfy 
both conditions (1) and (2) unless $n=2$, since 
any tree belonging to the diagrammatic expression of 
$r_n(\alpha, \phi^{\otimes n-1})$ for $n\geq 3$ must contain 
at least a branch of the two types depicted in figure 2. It follows 
that all terms in the sum of (\ref{gauge_W}) vanish except for 
the summand $n=2$. Since we clearly have $\alpha\bullet \phi\in K^1$
(by virtue of (\ref{der_dagger}) and (\ref{metric_invar})), it follows that
$r_2(\alpha,\phi)=
P(\alpha\bullet \phi)=\alpha\bullet\phi$, and thus 
$m_2(\alpha,\phi)=[\alpha,\phi]$. This shows that (\ref{gauge_W}) reduces 
to (\ref{moo}).

Relation (\ref{der_dagger}) implies that $K^0$ is an associative subalgebra
of $({\cal H}, \bullet)$, and hence ${\bf g}^0_W:=(K^0, [.,.])$ is a 
Lie subalgebra of string field gauge algebra
${\bf g}^0=({\cal H}^0, [.,.])$. In particular, the gauge algebra 
$g_W$ closes away from the critical set of $W$ and can be identified 
with the Lie algebra ${\bf g}_W^0$.
Transformation (\ref{moo}) integrates to
\be
\label{adjoint_W}
\phi\rightarrow \phi^\alpha=e^{ad_\alpha}\phi~~.
\ee
Hence equations (\ref{mc_W}) 
are invariant with respect to the 
symmetry group $G_W$ obtained by exponentiating the Lie 
algebra $g_W\equiv{\bf g}_W^0=(K^0,[.,.])$, and  
$\phi\in K^1$ transforms in its adjoint representation.

\subsection{Local equivalence of the two constructions}

What is the relation between ${\cal M}$ and ${\cal M}_W$ ? It is remarkable 
fact, which is discussed in more 
detail in Appendix A, that the two moduli spaces 
are isomorphic\footnote{More precisely, the two deformation functors are 
equivalent.}:
\be
{\cal M}_W\stackrel{\rm locally}{\approx}{\cal M}~~.
\ee
This follows from the observation \cite{Merkulov, Polishchuk, Merkulov_defs, 
Kontsevich_Soibelman}
that the algebras $(K, \{r_n\}_{n\geq 3})$ and 
$({\cal H}, Q, \bullet)$ are  {\em quasi-isomorphic} as $A_\infty$ 
algebras, i.e. their products are related by a sequence of maps which 
satisfy certain constraints and whose first element induces
an isomorphism between $K$ and the BRST cohomology of ${\cal H}$ 
(the precise definition is recalled in Appendix A). 
In fact, the homotopy algebra $(K, \{r_n\}_{n\geq 2})$ is a 
so-called {\em minimal model}  \cite{Keller} for $({\cal H}, Q, \bullet)$, 
if the later is viewed as an $A_\infty$ algebra whose third and higher 
products vanish. 

The quasi-isomorphism in question is defined by a sequence of maps 
$F_n:K\rightarrow {\cal H}$ obtained upon replacing $P$ with $U$ in 
the definition (\ref{rs}) of the string products $r_n$:
\be
\label{Fs}
r_n(u_1\dots u_n)=U\lambda_n(u_1\dots u_n)~~,
\ee
for $u_1\dots u_n\in K$. This defines $F_n$ for $n\geq 2$. One also needs 
a map $F_1:K\rightarrow {\cal H}$, which we take to be the inclusion
(this induces an isomorphism between $K$ and $H_Q({\cal H})$ by Hodge 
theory, which is why we obtain a quasi-isomorphism). 
This explicit construction of $F$ is due to  \cite{Kontsevich_Soibelman}.
 
Since $F$ gives a quasi-isomorphism of $A_\infty$ algebras, it is 
reasonable to expect that it also gives a quasi-isomorphism of $L_\infty$ 
algebras between their commutator algebras $(K, \{m_n\})$ and 
$({\cal H}, Q, [.,.])$. This somewhat elementary statement is proved 
in Appendix A\footnote{The statement is undoubtedly 
known to experts in homotopy algebras, but I include a proof 
since I could not find a convenient reference.}.

The final step is to recall from  \cite{Kontsevich_Felder} 
that the so-called deformation functors 
of two quasi-isomorphic $L_\infty$ algebras are equivalent, 
which implies that the associated moduli spaces are isomorphic.
In our case, the isomorphism follows by noticing that the map:
\be
\label{Fstar}
\phi\rightarrow F_*(\phi)=\sum_{n\geq 1}{F_n(\phi^{\otimes n})}
\ee
takes solutions of the extremum equations (\ref{mc_L}) into solutions 
of the string field equations of motion (\ref{mc}). The inverse correspondence
follows from the general result \cite{Kontsevich_Felder} that a 
quasi-isomorphism of 
$L_\infty$ algebras always admits a quasi-inverse, 
i.e. there exists an $L_\infty$ 
quasi-isomorphism 
$G:\{g_n:{\cal H}^n\rightarrow K\}_{n\geq 1}$ such that $G_1$ induces 
the inverse isomorphism $(G_1)^*=(F_1^*)^{-1}$ between $H_Q({\cal H})$ 
and $K$. Once such a quasi-inverse has been chosen, one obtains 
a map:
\be
\phi\rightarrow G_*(\phi)=\sum_{n\geq 1}{G_n(\phi^{\otimes n})}
\ee
which takes solutions of (\ref{mc}) into solutions of (\ref{mc_L}). Upon 
combining these two facts, it is not hard to prove the desired equivalence of 
deformation functors  \cite{Kontsevich_Felder}. We will have no need 
for this inverse correspondence, so we shall omit its explicit realization.

It follows that {\em one can compute the moduli space of a 
cubic string field theory either by solving the string field equations
of motion, or by extremizing the potential $W$, 
and the two results are locally assured to coincide}. 
Which of these two points of 
view one chooses depends on what is more convenient in the problem at hand. 
The cubic formulation gives the simpler-looking Maurer-Cartan equations 
(\ref{mc}), but requires knowledge of the BRST operator $Q$. 
The gauge-fixed formulation does not require this datum, but involves 
the entire sequence of products $r_k$.

\paragraph{Observations}

1. The correspondence (\ref{Fstar}) mixes the order of deformations. 
For example, if we have a solution $\phi=\sum_{i}{t_i\phi_i}$ 
to (\ref{mc_L}) in 
some deformation parameters $t_i$ (where $\phi_i$ form a basis of $K^1$), 
then the corresponding solution 
$F_*(\phi)$ of (\ref{mc}) involves higher orders in $t_i$. In terms of the 
associated moduli spaces, this means that (versal) solutions
to (\ref{mc_L}) and 
(\ref{mc}) describe local coordinate systems on ${\cal M}\approx{\cal M}_W$
which differ by a change of coordinates given by a power series.

2. Our explicit description of the potential gives a general method 
for computing this quantity. This description agrees manifestly with 
string perturbation theory.
 
3. Our potential depends on the choice of metric $h$ which enters 
the gauge-fixing procedure. 
It is clear that 
a change $h\rightarrow h'$ 
of this metric induces a quasi-isomorphism between the 
resulting $A_\infty$ algebras $(K, \{r_n\})$ and $(K', \{r'_n\})$ of string 
products. By the same argument as above, this implies  
that the resulting moduli spaces ${\cal M}_W$ and ${\cal M}_W'$ are 
isomorphic. Hence a change in the choice of metric corresponds to 
change of coordinates on ${\cal M}$. The associated transition functions 
will generally involve power series.

4. That the two descriptions of the moduli space agree was 
expected based on the physical interpretation of $W$ as 
a tree-level potential for the physical modes. The fact that this 
intuitive interpretation is strictly correct is, however, entirely 
nontrivial. As we saw above, its proof makes heavy use of results 
in modern deformation theory.

5. Our construction gives a string-field theoretic 
explanation for the appearance of homotopy algebras in {\em cubic} 
string field theory. Its application to the topological A and B models 
(to be discussed below) gives one reason for 
the relevance of such structures in homological 
mirror symmetry  \cite{Kontsevich}. It is a general principle of modern 
topology and deformation theory that many problems can be better 
understood by enlarging the class of differential graded 
(associative, Lie\dots ) 
algebras to the class of their homotopy versions. The double description of 
moduli~spaces discussed above gives an explicit example of the relevance 
of this principle to string field theory. The fact that homotopy structures 
play a fundamental role in string theory can be traced back to its relation 
with loop spaces  \cite{Sullivan}\footnote{I am grateful to D. Sullivan for 
an illuminating discussion of these issues.}. 
It is clear for many reasons that 
a deeper understanding of string theory requires systematic use of this 
language. For work in this direction I refer 
the reader to the basic references  \cite{Zwiebach_Witten, Zwiebach_closed}, 
as well as to the more recent  papers  \cite{Gaberdiel, 
Zwiebach_open, Hofman1, Hofman2, JS}. 

\subsection{An intrinsic formulation of D-brane moduli spaces}

In physical applications, cubic string field theory arises 
as an off-shell description of the dynamics of open 
strings whose endpoints lie on a D-brane (Figure 3). 
(This includes the case of purely von Neumann boundary conditions).
In this 
situation, one is interested in giving a string-theoretic 
description of the associated D-brane moduli space.

\hskip 1.0 in
\begin{center} 
\scalebox{0.5}{\begin{picture}(0,0)%
\includegraphics{brane.pstex}%
\end{picture}%
\setlength{\unitlength}{4144sp}%
\begingroup\makeatletter\ifx\SetFigFont\undefined%
\gdef\SetFigFont#1#2#3#4#5{%
  \reset@font\fontsize{#1}{#2pt}%
  \fontfamily{#3}\fontseries{#4}\fontshape{#5}%
  \selectfont}%
\fi\endgroup%
\begin{picture}(3690,1838)(406,-1700)
\put(2251,-871){\makebox(0,0)[lb]{\smash{\SetFigFont{17}{20.4}{\familydefault}{\mddefault}{\updefault}
$a$
}}}
\put(406,-511){\makebox(0,0)[lb]{\smash{\SetFigFont{17}{20.4}{\familydefault}{\mddefault}{\updefault}
${\cal H}$
}}}
\put(4096,-1051){\makebox(0,0)[lb]{\smash{\SetFigFont{17}{20.4}{\familydefault}{\mddefault}{\updefault}
${\cal M}_a\equiv {\cal M}_{\cal H}$
}}}
\end{picture}
}
\end{center}
\begin{center} 
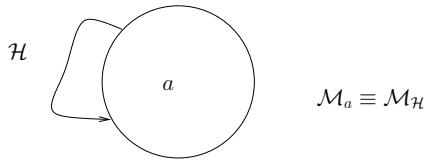
Figure  3. {\footnotesize We define the moduli space of a D-brane $a$ 
to be 
the moduli space of the open string field theory of strings stretching from 
$a$ to $a$. The latter is defined on the off-shell 
state space ${\cal H}$ of such strings.} 

\end{center}

When studying such 
moduli spaces, the prevalent procedure has been to approach the problem 
either from a space-time, geometric point of view, or from a 
sigma~model or conformal field theory perspective. In the first case, one 
identifies D-brane moduli with the moduli of some geometric object, and 
then attempts to compute its `stringy corrections' by considering some 
auxiliary construction (such as the partially-wrapped 
D-branes of  \cite{Douglas_quintic, Kachru1}). 
The second approach  \cite{Aspinwall}
relies on a study of marginal deformations of a boundary conformal field 
theory  \cite{bcft_defs}. 

I would like to propose a different perspective on this issue, which 
attacks the problem via the methods of 
string field theory. Namely, {\em we define deformations of the 
D-brane $a$ to be the vacuum deformations of the string field 
theory of open strings whose endpoints end on $a$}. If the latter admits 
a cubic formulation, then the resulting moduli space can be described 
in terms of the Maurer-Cartan equation (\ref{mc})\footnote{This 
perspective is the starting point of the 
papers  \cite{com1, com3} which gave a general analysis of moduli spaces 
and condensation processes in a system containing an arbitrary collection of 
D-branes.}.
It then follows from our results that the same space can be 
described in terms of the extremum equations (\ref{mc_L}) for the 
tree level potential $W$. This gives an alternate 
formulation of the same problem, which is equivalent with 
the string field theory approach, and recovers some of 
the conformal field theory perspective in a computationally efficient manner.
In the next section, we apply this method to the 
simple case of open topological A/B strings compactified on 
a Calabi-Yau threefold.

\section{Application to topological A/B models}

\subsection{The B model}

\subsubsection{The geometric data}

We consider a Calabi-Yau threefold $X$ and a B-type brane described 
by a 
holomorphic vector bundle $E$ over $X$. The string field theory 
of strings whose endpoints lie on this brane 
is the (open) holomorphic Chern-Simons theory of  \cite{Witten_CS}. 
This has the off-shell state space
${\cal H}=\Omega^{0,*}(E^*\otimes E)$, the associative product 
$\bullet=\wedge$ 
given by wedge product of bundle-valued forms (this includes composition of 
bundle morphisms), and the BRST operator $Q_E={\overline \partial}$
(the Dolbeault differential coupled to the bundle $E$). 
The worldsheet degree is given by form rank ($|u|=p$ if 
$u\in \Omega^{0,p}(E^*\otimes E)$) and the bilinear form
on ${\cal H}$ is given by:
\be
\label{B_metric}
\langle u, v \rangle=\int_{X}{\Omega\wedge tr_E(u\wedge v)}~~,
\ee
where $tr_E$ denotes the fiberwise trace on the bundle $End(E)=E^*\otimes E$.

In order to obtain 
a perturbative expansion, one must choose a gauge and define propagators. 
To this end,  we pick a Hermitian metric $g_E$ on $E$. Together 
with the Calabi-Yau metric $g$ on $X$, this 
induces a metric $(.,.)_E$ on the bundle $\Lambda^*(T^*X\oplus 
{\overline T}^*X)
\otimes End(E)$. 
If $u=\omega\otimes\alpha$ and $v=\eta\otimes \beta$ are decomposable 
elements of $\Omega^{*,*}(End(E))$, then: 
\be
(u,v)_E=(\omega,\eta)tr_E(\alpha^\dagger\circ\beta)~~,
\ee
where $(\omega,\eta)$ is the metric induced by $g$ on 
$\Lambda^*(T^*X\oplus {\overline T}^*X)$, 
normalized as in \cite{GH}. Note that we take all Hermitian metrics
to be antilinear with respect to the {\em first} variable. 
The metrics $(.,.)$ and $(.,.)_E$ allow one to define 
antilinear Hodge operators ${\overline *}$ and ${\overline *}_E$ on 
$\Omega^{*,*}(X)$ and $\Omega^{*,*}(E^*\otimes E)$,
which take $\Omega^{p,q}(X)$ into $\Omega^{3-p,3-q}(X)$ and
$\Omega^{p,q}(E^*\otimes E)$ into 
$\Omega^{3-p,3-q}(E^*\otimes E)$. The operator ${\overline *}_E$ is
the tensor product of ${\overline *}$ with the Hermitian conjugation 
$\dagger$ in 
$End(E)$, taken with respect to the metric $g_E$:
\be
{\overline *}_E(\omega\otimes \alpha)={\overline *}\omega \otimes 
\alpha^\dagger~~.
\ee
With these conventions, one has the relations:
\be
\label{rels}
{\overline *}1_X=vol_g~~,~~
({\overline *}\omega)\wedge \eta=(\omega, \eta)vol_g~~,~~
tr_E[({\overline *}_Eu)\wedge v]=(u,v)_E vol_g~~,
\ee
where $1_X$ is the unit function defined on $X$, $1_E$ is the identity 
section of $End(E)$ and $vol_g$ is the volume form induced by the Calabi-Yau 
metric $g$. We recall that ${\overline *}_E$  
satisfies:
\be
\label{square}
{\overline *}_E^2u=(-1)^{rk u}u~~,
\ee
as a consequence of the similar property of ${\overline *}$. 

\subsubsection{The operator $c$}

We next define a Hermitian product $h$ on ${\cal H}$ 
through the standard relation:
\be
\label{Hermitian}
h(u,v)=\int_{X}{tr_E[({\overline *}_Eu)\wedge v]}=
\int_{X}{(u, v)_E vol_g}~~.
\ee
where $u,v\in \Omega^{0,*}(E^*\otimes E)$. 
This scalar product vanishes on homogeneous elements $u,v$ unless 
their ranks coincide. The relation between ${\overline *}_E$ and
$h$ is similar to the relation between $c$ and $h$ discussed in Section 2, 
but with respect to the `wrong' bilinear form 
$\xi(u,v)=\int_{X}{tr_E(u\wedge v)}$ (rather than the physically correct 
form (\ref{B_metric})). In fact, the Hodge operator ${\overline *}_E$ 
does not preserve the space 
${\cal H}$, and therefore it cannot be identified with the conjugation $c$.
Indeed, ${\overline *}$ maps $\Omega^{0,q}(E^*\otimes E)$ into 
$\Omega^{3,3-q}(E^*\otimes E)$, which is not a subspace of ${\cal H}$. 
Following the discussion of Section 2, we define an operator 
$c:\Omega^{0,*}(E^*\otimes E)\rightarrow \Omega^{0,3-*}(E^*\otimes E)$
through the relation $h(u,v)=\langle cu,v\rangle$, i.e.:
\be
\int_{X}{tr_{E}\left[({\overline *}_E u) \wedge v\right]}=
\int_{X}{\Omega\wedge tr_E(cu\wedge v)}~~.
\ee
Since this must hold for all $v$, we take:
\be
\label{c_def}
{\overline *}_E u=\Omega\wedge cu~~,
\ee
which uniquely determines the antilinear map $c$. In order to satisfy the 
framework of Section 2, we must check that $c$ squares to the identity. 
We show that this 
can be fulfilled by normalizing the holomorphic 3-form through:
\be
\label{norm1}
({\overline *}\Omega)\wedge \Omega=vol_g~~,
\ee 
where $vol_g$ is the volume form induced by the Calabi-Yau metric $g$.  
This normalization condition can always be satisfied by 
a constant rescaling of $\Omega$. The proof of the identity $c^2=id$ 
proceeds in three steps: 

\

\noindent (1) We first show the relation:
\be
\label{csquare}
tr_E(c^2u \wedge v)=tr_E(u\wedge c^2v)~~, {\rm~for~}
u,v\in {\cal H}=\Omega^{0,*}(End(E))~~.
\ee
For this, notice that the last of equations (\ref{rels}) and the 
definition (\ref{c_def}) imply:
\be
\Omega\wedge tr_E(cu\wedge v)=(u,v)_Evol_g~~.
\ee
Upon permuting $u$ and $v$ in this equation and using the  
hermicity of $(.,.)_E$, one obtains:
\be
\label{sq}
\Omega\wedge tr_E(cu\wedge v)=\overline{\Omega\wedge tr_E(cv\wedge u)}~~.
\ee 
Relation (\ref{csquare}) now follows by repeated application of (\ref{sq}), 
combined with the graded symmetry of the wedge product. 

\

(2)\noindent We next compute the value of $c^2(\alpha)$ for $\alpha$ a 
local section 
of $End(E)$. According to (\ref{c_def}), the 
$(0,3)$ form $c(\alpha)$ is determined by:
\be
\label{c_d}
{\overline *}_E \alpha=\Omega\wedge c(\alpha)~~.
\ee
On the other hand, the first 
relation in (\ref{rels}) implies 
${\overline *}_E \alpha=vol_g \otimes \alpha^\dagger$. 
Combining this with (\ref{c_d}) and (\ref{norm1}) gives:
\be
c(\alpha)=-{\overline *}\Omega\otimes \alpha^\dagger=
-{\overline *}_E(\Omega\otimes \alpha)~~.
\ee
We next determine the 0-form 
$c^2(\alpha)=-c({\overline *}\Omega\otimes \alpha^\dagger)$ 
from the defining relation 
(\ref{c_def}):
\be\label{csq}
{\overline *}_E({\overline *}\Omega\otimes \alpha^\dagger)=
\Omega\wedge c({\overline *}\Omega\otimes \alpha^\dagger)~~
\Leftrightarrow \Omega\otimes \alpha =
-\Omega \otimes c({\overline *}\Omega\otimes \alpha^\dagger)\Leftrightarrow 
-c({\overline *}\Omega\otimes \alpha^\dagger)=\alpha~~,
\ee
where we used ${\overline *}_E({\overline *}\Omega\otimes \alpha^\dagger)=
{\overline *}_E^2(\Omega\otimes \alpha)=-\Omega\otimes \alpha$, 
due to property (\ref{square}) of the Hodge operator. 
Equation (\ref{csq}) shows that:
\be
\label{c_sq1}
c^2(\alpha)=\alpha~~.
\ee
This nice relation is a consequence of the normalization condition 
(\ref{norm1}). 

\

(3)\noindent We are now ready to show that $c^2=Id_{\cal H}$. 
For this, we take $v=\alpha$ in relation (\ref{csquare}) and use the 
property (\ref{c_sq1}) to obtain:
\be
tr_E[(c^2u)\alpha]=tr_E(u\alpha)~~.
\ee
Since $\alpha$ is an arbitrary local section of $End(E)$, this 
leads to the desired conclusion: 
\be
c^2(u)=u~~{\rm~for~}u\in {\cal H}=\Omega^{0,*}(End(E))~~.
\ee
Hence all arguments of Sections 2 and 3 apply. 

\subsubsection{The potential}

Proceeding as in Section 2, 
we pick the gauge\footnote{This is an analogue of the 
Siegel gauge of bosonic string theory. Indeed, the analogue of the 
bosonic antighost operator $b_0$ in a topological string theory obtained 
by twisting an $N=2$ superconformal field theory is the generator 
${\cal G}_0$ of the topological (`twisted') $N=2$ algebra. In a unitary 
conformal field theory, this is the Hermitian conjugate of the generator 
${\cal Q}_0$, which can be identified with the BRST charge. In our treatment, 
this is represented in spacetime through the Dolbeault differential 
${\overline \partial}$ via the localization argument of  \cite{Witten_CS}.}:
\be
Q_{E}^+u=0~~{\rm~for~} u\in \Omega^{0,*}(End(E))~~,
\ee
where $Q_{E}^+={\overline \partial}^+=-{\overline *}{\overline \partial} 
{\overline *}$ is the Hermitian conjugate of $Q_{E}={\overline \partial}$ 
with respect to the scalar product (\ref{Hermitian}) (this can also 
be expressed in the form (\ref{qd})). 
The associated Hamiltonian $H=QQ^++Q^+Q$ is the 
${\overline \partial}$-Laplacian:
\be
H=\Delta_{\overline \partial}={\overline \partial}^+{\overline \partial}+
{\overline \partial}{\overline \partial}^+~~.
\ee
The BRST cohomology is then represented by physical states, 
i.e. degree one states lying in the kernel $K_{E}=ker Q_{E}\cap ker Q_{E}^+$; 
these are just the harmonic forms in $\Omega^{0,1}(End(E))$. The operator 
$P$ of Section 2 is the orthogonal projector on the space $K$ of 
harmonic $(0,*)$-forms, while the propagator $U$ has the form:
\be
U={\overline \partial}^+G=\frac{1}{\Delta_{\overline \partial}}
{\overline \partial}^+~~,
\ee
where $G=\frac{1}{\Delta_{\overline \partial}}(1-P)$ is a Green's function 
for $\Delta_{\overline \partial}$. 

As in Section 2, one can express the 
disk string correlators $\langle \langle u_0\dots u_n\rangle \rangle^{(n)}$ 
of $n+1$  physical states in terms of tree level Feynman rules.
The resulting potential has the form:
\be
W(\phi)=\sum_{n\geq 2}{\frac{(-1)^{n(n+1)/2}}{n+1}
\int_{X}{\Omega\wedge tr_E(\phi\wedge r_n(\phi^{\otimes n}))}}~~,
\ee
with $\phi \in \Omega^{0,1}_{{\overline \partial}-harm}(End(E))$ and 
the products $r_n$ defined as explained there. It is easy to see that these 
coincide with the products considered in  \cite{Polishchuk}, 
were they were introduced without a physical justification.
As explained there, $r_n$ induce the 
holomorphic version of Massey products on 
$H^{0,*}_{\overline \partial}(End(E))\approx H^*_{sheaf}(End(E))$, 
when the latter are well-defined and one-valued. Unlike the Massey products,
though, the string products $r_n$ are {\em always} well-defined; 
hence they give an extension of classical Massey theory.  

The argument of Section 2 implies that $r_n$ have the cyclicity properties 
(\ref{cycl}) {\em with respect to the topological metric} (\ref{B_metric}).
I wish to caution the reader that this is {\em not} the bilinear form used in 
 \cite{Polishchuk}. Indeed, the paper cited proves cyclicity with respect 
to the `wrong' bilinear form $\xi(u,v)=\int_{X}{tr_E(u\wedge v)}$, 
which is related to Serre duality. As explained above, this does  
not coincide with the bilinear form relevant for B-model physics. 
The physically relevant object is the topological metric (\ref{B_metric}), 
and our construction of the operator $c$, together with the general arguments 
of Section 2, show that the cyclicity result of  \cite{Polishchuk} remains 
valid with respect to this form.

It is clear that $W(\phi)$ corresponds to a particular case of 
the `D-brane superpotentials'
considered in  \cite{Douglas_quintic, Kachru1}. In our situation, 
this interpretation arises 
upon considering some $9$-brane partially wrapped over $X$ and filling 
the uncompactified 
spacetime directions. As mentioned above, this point of view 
is not intrinsic and is affected by difficulties 
due to flux conservation (which must be cured by restricting to 
non-compact Calabi-Yau manifolds or positing some 
orientifold construction). For these reasons, 
we prefer the string field interpretation.

\subsubsection{The moduli space}

As outlined above, it is natural to define the moduli space 
${\cal M}_E$ of our B-type brane
as the moduli space of vacua 
of the associated string field theory. As discussed 
in Section 3, this admits two equivalent presentations: 

\paragraph{String field theory description}
In this approach, ${\cal M}_E$ is obtained by solving the 
Maurer-Cartan equations:
\be
{\overline \partial}\phi+\phi\wedge \phi=0~~
\ee  
and dividing through the gauge group action generated by:
\be
\phi\rightarrow \phi-{\overline \partial}\alpha-[\phi, \alpha]~~,
\ee
with $\alpha\in \Omega^{0,0}(End(E))=\Gamma(End(E))$ a 
(smooth, i.e. ${\cal C}^\infty$) section of $End(E)$. 
The Maurer-Cartan equation is equivalent with 
$({\overline \partial}+\phi)^2=0$, which is the condition that 
${\overline \partial}+\phi$ determines an integrable complex structure;
the gauge transformations identify equivalent complex structures. 
The string field theory gauge group ${\cal G}$ can be identified
with the group $Aut(E)$ of smooth (${\cal C}^\infty$)
automorphisms $\lambda$ of $E$, whose adjoint action on 
${\cal H}=\Omega^{0,*}(End(E))$ has the form:
\be
u\rightarrow \lambda \circ u \circ \lambda^{-1}~~.
\ee
Finite gauge transformations of the string field
$\phi \in {\cal H}^1=\Omega^{0,1}(End(E))$ are given by:
\be
\phi\rightarrow \lambda \circ {\overline \partial}\lambda^{-1} 
+ \lambda\circ \phi \circ \lambda^{-1}~~.
\ee
\noindent Therefore, ${\cal M}_E$ is the moduli space of complex 
structures on $E$ (more precisely, its component which 
contains the original complex structure ${\overline \partial}$). 
Note that both the space of integrable connections 
and the gauge group are infinite dimensional, so we have an `infinite
presentation' of the finite-dimensional space 
${\cal M}_E$, i.e. a presentation as the quotient of an infinite-dimensional
space through the action of an infinite dimensional transformation group.

\paragraph{Description trough the D-brane superpotential}
According to our general arguments,  
the same moduli space can be described in terms of 
solutions to the `F-flatness' condition $\frac{\delta W}{\delta \phi}=0$,  
which leads to equation (\ref{mc_W}). To identify the symmetry group 
$G_W$, notice that $K^0$ is the space $\Gamma_{hol}(End(E))$ of 
(global) holomorphic 
sections of $End(E)$. It is clear that 
${\overline \partial}^+(\alpha u)=\alpha{\overline \partial}^+u$ 
for any holomorphic section $\alpha$ of $End(E)$ and any $End(E)$-valued 
$(0,q)$-form $u$; thus hypothesis (\ref{der_dagger}) of 
Subsection 3.3. is satisfied.  The Lie algebra of $G_W$ is 
$(K^0, [.,.])$, which is simply the Lie algebra $\Gamma_{hol}(End(E))$\
of (global) holomorphic sections of $End(E)$, while the effective 
symmetry group is the group $Aut_{hol}(E)$ of holomorphic automorphisms 
of $E$. This acts on 
${\cal H}^1=\Omega^{0,1}(End(E))$ in its adjoint representation: 
\be
\label{G_WB}
\phi\rightarrow \lambda\phi \lambda^{-1}~~,
\ee
for $\phi\in \Omega^{0,1}(End(E))$ and $\lambda\in Aut_{hol}(E)$. 
This action preserves the space $K^1$ of $End(E)$-valued harmonic 
$(0,1)$-forms, on which it restricts to the action
(\ref{adjoint_W}) of Subsection 3.3.
Hence the moduli space of complex structures on $E$ can be locally 
described as the space of solutions 
$\phi\in K^1=\Omega^{0,1}_{{\overline \partial}-harm}(End(E))$ 
of the `F-flatness conditions' 
$\frac{\delta W}{\delta \phi}=0$, modded out by the group action (\ref{G_WB}).
This gives the precise realization for our system of the  
proposals of  \cite{Douglas_quintic} and  \cite{Kachru1}. The result is  
a `finite presentation' of ${\cal M}_E$, as the quotient of a 
finite-dimensional space through the action of a finite-dimensional 
group.

\paragraph{Observation} A thorough analysis of the two presentations
involves functional analytic, respectively 
complex-analytic/algebro-geometric
issues. How to deal with these is a standard
subject. Our description of the moduli problems 
is in fact entirely formal (even locally), since we did not 
address the problem of building a complex analytic structure 
on the moduli space (which requires Kuranishi theory).
Strictly speaking, we are only studying 
the associated deformation functors, rather than the moduli spaces 
themselves.

\subsection{The A model}

The $A$ model realization is also easy to obtain (part of 
this is already sketched in  \cite{Witten_CS}). For this, we consider an 
$A$-type brane described by a pair $(L, E)$ with $L$ a special Lagrangian 
cycle of a Calabi-Yau threefold $X$ and $E$ a flat complex vector 
bundle over $E$. 
We remind the reader that a flat structure on $E$ can be described either in 
terms of a collection of local trivializations whose transition functions 
are constant, or in terms of a flat connection $A$ on $E$. The latter 
defines a differential $d$ on $End(E)$-valued forms, the Dolbeault 
differential `twisted by $A$'. The string field theory in our boundary 
sector is described by the data ${\cal H}=\Omega^*(E)$, $Q=d$, 
$\bullet=\wedge$ and:
\be
\langle u, v \rangle=\int_{L}tr_E(u\wedge v)~~{\rm~for~}u, v \in 
\Omega^*(L, End(E))~~.
\ee
This is the Chern-Simons field theory on $E$:
\be
S(\phi)=\int_{L}{tr_E\left[\frac{1}{2}\phi \wedge d\phi+\frac{1}{3}
\phi\wedge \phi\wedge \phi\right]}~~.
\ee
The perturbation expansion is obtained by picking a Hermitian metric 
on $E$. If $*$ is 
the associated Hodge operator, we have an induced Hermitian 
product on ${\cal H}$ given by:
\be
h(u,v)=\int_{L}{tr_E(*u\wedge v)}~~.
\ee
The Hermitian conjugate of $Q=d$ with respect to this product is 
$Q^+=d^+=-*d*$, and the associated Hamiltonian is the $d$-Laplacian, 
$H=\Delta=dd^++d^+d$ (coupled to the flat bundle $E$). 
The space $K^1$ of physical states consists of $d$-harmonic 
$End(E)$-valued one-forms on $L$. In this case, one has $*^2=Id$ and 
the antilinear map 
$c$ of Section 3 is simply the Hodge operator $*$. The string field 
theory gauge group ${\cal G}$ is the group of ${\cal C}^\infty$ 
automorphisms of $E$, acting on string fields 
$\phi\in {\cal H}^1=\Omega^1(End(E))$ through:
\be
\phi\rightarrow \lambda\circ \phi\circ \lambda^{-1}+\lambda \circ d\lambda^{-1}~~.
\ee
It is clear that the Maurer-Cartan equations (\ref{mc}) give the moduli 
space ${\cal M}_E$ of flat connections on $E$. 
This can also be described 
in terms of extrema of the potential:
\be
W(\phi)=\sum_{n\geq 2}{\frac{(-1)^{n(n+1)/2}}{n+1}
\int_{L}{tr_E(\phi\wedge r_n(\phi^{\otimes n}))}}~~,
\ee
with the string products $r_n$ built as in Section 3. The effective 
symmetry group
$G_W$ is the group $Aut_{flat}(E)$ 
of covariantly constant gauge transformations
(automorphisms of $E$ as a {\em flat} vector bundle). This acts on 
elements of $K^1=\Omega^1_{harm}(End(E))$ 
through its adjoint representation. 
We obtain a finite presentation of the moduli space of flat connections 
on $E$.

Since we wish to obtain a {\em complex} moduli space, 
{\em we do not require 
the one-forms $\phi$ to be anti-Hermitian}. This amounts to considering flat 
connections on $E$ which are not subject to a hermicity condition 
(hence the 
Chern-Simons action becomes complex-valued, as was the case for the B-model; 
alternately, one can replace this complex action through its imaginary part
etc).
Similarly, elements of the groups ${\cal G}$ and $G_W$ are not required to
be unitary.
The (somewhat non-rigorous) justification for this is that the moduli space 
of flat anti-Hermitian connections  
describes only half of A-brane moduli. The other half, which 
is due to deformations of the special Lagrangian cycle, is expected
to pair up with the former due to $N=1$ space-time supersymmetry -- a 
fact which is locally described by lifting the antihermicity condition. 
This can be seen explicitly for the case of a singly-wrapped brane 
(i.e. $rank E=1$), upon using the results of  \cite{McLean} (see 
 \cite{Hitchin} for a nice exposition). For 
multiply-wrapped branes ($rank E>1$), the situation is less clear, 
though it seems  \cite{Zaslow_recent} that one could study that case 
by viewing such objects as degenerations of singly-wrapped 
branes. The procedure of lifting the antihermicity 
constraint should be trusted only locally -- the global structure of the 
moduli space is clearly much more involved.  A thorough physical analysis of 
this issue seems to require an extension of the open A-model of 
 \cite{Witten_CS}, which 
should be obtained by considering supplementary boundary couplings
(couplings to sections of the normal bundle to the special Lagrangian cycle 
$L$, which describe its deformations). This sort of extension does not seem 
to have been studied, despite its relevance for 
mirror symmetry and Calabi-Yau D-brane physics\footnote{The results of 
 \cite{McLean} assure that deformations of a special Lagrangian cycle $L$ 
can be represented through one-forms on the cycle. In the presence of 
a multiply-wrapped brane on $L$, these do not pair up with deformations 
of the flat connection in any obvious manner. One way out is to view 
a multiply wrapped brane as involving $n$ copies of the cycle 
$L$ lying on top of each other, and construct the higher rank 
Chan-Paton bundle $E$ on $L$ from a collection of 
line bundles, each  living on one of the copies of $L$. This is complicated by 
the fact that one generally must also take the B-field into account.
It is 
unclear if such a construction allows one to recover the most general 
physical situation. 
Moreover, its significance is doubtful unless one can show
equivalence with a generalized A model which explicitly couples to 
deformations of $L$.}. 
An analysis of such a generalized model would presumably make 
contact with the mathematical construction of  \cite{Merkulov_sl}.

\section{Conclusions and directions for further research}

We presented a detailed analysis of tree level boundary potentials 
in (cubic) open string field theory, giving a general prescription for 
their construction in terms of the associated 
string products. By analyzing the resulting moduli problem, we gave 
its formulation in terms of modern deformation theory and 
{\em proved} the equivalence of the string field and `low energy' 
descriptions of the moduli space. The proof makes 
use of recent results in the theory of homotopy algebras. 
Upon applying these methods to the topological A and B models, we 
identified our potentials with the D-brane superpotentials of 
 \cite{Douglas_quintic, Kachru1} and gave 
their precise description in terms of geometric 
quantities. This clarifies the meaning of these objects 
and explains the appearance of homotopy associative and homotopy Lie algebras 
in the {\em associative} open string field theories of the B-model and in 
the large radius limit of the A-model, thereby establishing part of the 
connection with mathematical work on homological mirror symmetry.
The fact that this connection appears naturally from the string field 
point of view brings further evidence in favor of the program 
outlined in  \cite{com3} of recovering (and potentially extending) 
homological mirror symmetry by means of string field theory.

The fact that deformations of 
flat and holomorphic vector bundles are potential in this sense 
is a direct consequence of the existence of a string field theory 
description. This can be formulated more generally in deformation-theoretic 
terms, by introducing a bilinear form and appropriate `conjugation' operator
$c$ within the framework of classical (Maurer-Cartan) or generalized 
(homotopy Maurer-Cartan) deformation theory (with a non-vanishing first 
order product). The first situation corresponds 
to cubic (or associative) open string field theory  \cite{Witten_CS}, 
while the second is described
by the non-polynomial (homotopy associative) open string field theory of 
 \cite{Gaberdiel, Zwiebach_open}. 
The arguments of the present paper can be generalized 
to the second case, which implies the existence of a potential for 
a variety of deformation problems important in physics. For example, 
it seems that this approach can shed light on the existence of a potential 
for holomorphic curves embedded in a Calabi-Yau threefold, thereby 
explaining and generalizing some results of  \cite{Kachru1, Katz}. We hope 
to return this issue in future work. 

Another extension involves the inclusion of instanton 
corrections in the A model. One reason for restricting to the large 
radius limit is the fact that disk instantons are responsible for a series 
of effects which require a rather sophisticated analysis. 
A precise discussion can be given with string field theory methods, and 
makes contact with the work of K. Fukaya \cite{Fukaya, Fukaya2}.

Finally, it is important to extend the analysis of the present paper 
by including interactions between open and closed strings. This can be carried 
out by considering the tree-level restriction of the string field theory of 
 \cite{Zwiebach_open}, which is governed by a so-called $G_\infty$ 
(homotopy Gerstenhaber) algebra and provides the off-shell extension 
of the framework of  \cite{top}. The associated moduli space describes 
joint deformations of the Calabi-Yau manifold $X$ and a D-brane, for example
simultaneous deformations of the complex structure of $X$ 
and of a holomorphic vector bundle $E$ on $X$ 
(for the case of the B-model). The string field
approach should lead to an explicit description of the relevant 
potentials, as well a detailed analysis of the deformation problem (and, for 
the A-model, of its `quantization'). The 
infinitesimal version of such deformations was analyzed to first
order in  \cite{Hofman1}. As in the open string case, string field theory 
can provide a more explicit construction.

\acknowledgments{
I wish to thank S.~Popescu for collaboration in a related 
project and D.~Sullivan and E.~Diaconescu for useful conversations.
I also thank R. Roiban for comments on the manuscript.
I am endebted to M.~Rocek for constant support and interest in my work. 
The author is supported by the Research Foundation under NSF grant 
PHY-9722101. }

\

\appendix

\section{Homotopy algebras and their morphisms}

This appendix summarizes some basic facts about (strong) homotopy associative
and homotopy Lie algebras, and gives a proof of the statement that 
an $A_\infty$ quasi-isomorphism induces an $L_\infty$ quasi-isomorphism 
between  the commutator algebras. Most of the appendix is written 
in an explicit, 
`component' language. However, the proof itself makes use of a 
dual description in terms of codifferential coalgebras, since this 
avoids computational morass. For this, 
I shall assume that the reader is familiar with the 
theory of coalgebras. This appendix is intended for the convenience of 
non-expert readers. Readers familiar with the subject may 
wish to skip to Appendix B.

\subsection{$A_\infty$ algebras}
For more information on 
this subject the reader is referred to \cite{Merkulov_infty}, 
\cite{Lada_Markl}, \cite{Keller}, \cite{Proute} and 
\cite{Polishchuk_ell2}
and the references therein. A basic reference 
discussing the role of $A_\infty$ algebras in open string field theory is 
 \cite{Gaberdiel}.

\paragraph{Definition}
A (strict, or strong) 
$A_\infty$ algebra is defined by a $\Z$-graded vector space $A$
together with a countable collection of operations 
$r_n:A^{\otimes n}\rightarrow A$ $(n\geq 1$)
which are homogeneous of degree $2-n$ and
subject to the constraints:
\be
\label{ainf}
\sum_{\tiny \begin{array}{c}k+l=n+1\\j=0\dots k-1\end{array}}
{(-1)^s r_k(u_1\dots u_j,r_l(u_{j+1}\dots u_{j+l}),u_{j+l+1}\dots u_n)}=0~~,
\ee
for $n\geq 1$, where $s=l(|u_1|+\dots |u_j|)+j(l-1)+(k-1)l$ and $|~.~|$ 
denotes the degree of homogeneous elements in $A$.

\paragraph{Observation} A 
{\em weak} $A_\infty$ algebra  \cite{Fukaya, Fukaya2} is defined in a similar 
manner, but it also contains a $0^{th}$ order product $r_0$. We shall have no 
use for weak $A_\infty$ algebras in this paper. Such structures 
are relevant for 
describing open string field theory built around a background which does not 
satisfy the string equations of motion.

\

The first three constraints (\ref{ainf}) read:
\bea
\label{a3}
& &r_1^2=0~~\nn\\
& &r_1(r_2(u_1,u_2))=r_2(r_1(u_1),u_2)+(-1)^{|u_1|}r_2(u_1,r_1(u_2))~~\\
& & r_2(u_1, r_2(u_2, u_3))-r_2(r_2(u_1, u_2), u_3)=\nn\\
& &~~~~~~~
r_1(r_3(u_1,u_2,u_3))+
r_3(r_1(u_1), u_2, u_3)+(-1)^{|u_1|}r_3(u_1, r_1(u_2), u_3)
+\nn\\
& &~~~~~~~(-1)^{|u_1|+|u_2|}r_3(u_1, u_2, r_1(u_3))~~.\nn
\eea
In particular, $r_1$ is a degree one differential on $A$. In the case 
$r_n=0$ for $n\geq 3$, an $A_\infty$ algebra reduces to a differential graded 
associative algebra with differential $Q=m_1$ and product $u\bullet v=r_2(u,v)$.

\paragraph{Definition} Given a (strict) $A_\infty$ algebra 
$(A, \{r_n\}_{n\geq 1})$, its {\em cohomology} $H_{r_1}(A)$ 
is the cohomology of 
the vector space $A$  with respect to the differential $r_1$.

\paragraph{Definition \cite{Polishchuk_ell2}}
Given two (strict) $A_\infty$ algebras 
$(A, \{r_n\}_{n\geq 1})$ and $(A', \{r'_n\}_{n\geq 1})$, an $A_\infty$ 
{\em morphism} $f=\{f_n\}_{n\geq 1}$ from $A$ to $A'$ is a collection of 
maps $f_n:A^{\otimes n}\rightarrow A'$ which are homogeneous of degree $1-n$ 
and satisfy the conditions:
\bea
\label{amorphism}
\sum_{1\leq k_1<k_1\dots <k_i=n}{(-1)^{r}r'_i(f_{k_1}(u_1\dots u_{k_1}),
f_{k_2-k_1}(u_{k_1+1}\dots u_{k_2})\dots f_{n-k_{i-1}}(u_{k_{i-1}+1}\dots u_n))}=\nn\\
\sum_{k+l=n+1}\sum_{j=0}^{k-1}{(-1)^{|u_1|+\dots +|u_j|+j+s}
f_k(u_1\dots u_j,r_l(u_{j+1}\dots u_{j+l}), u_{j+l+1}\dots u_n)}~~,~~~~~~~~~
\eea
for $n\geq 1$. The exponents $r$ and $s$ in these relations are given by:
\bea
r&=&\mu(u_1,\dots ,u_{k_1})+\mu(u_{k_1+1},\dots ,u_{k_2})+\dots 
+\mu(u_{k_{i-1}+1},\dots ,u_n)+\nn\\
& &\mu(f_{k_1}(u_1,\dots ,u_{k_1}),
\dots ,f_{n-k_{i-1}}(u_{k_{i-1}+1},\dots ,u_n))~~,\\
s&=&\mu(u_{j+1},\dots ,u_{j+l})+
\mu(u_1,\dots ,u_{j},r_l(u_{j+1},\dots ,u_{j+l}),u_{j+l+1},\dots ,u_n)~~,\nn
\eea
where 
\be
\mu(a_1,\dots ,a_k):=(k-1)|a_1|+(k-2)|a_2|+\dots  +|a_{k-1}|+
\frac{k(k-1)}{2}~~.
\ee

The first two constraints in (\ref{amorphism}) read:
\bea
r'_1(f_1(u))&=&f_1(r_1(u))~~\\
r'_2(f_1(u_1),f_1(u_2))&=&
f_1(r_2(u_1,u_2))+r'_1(f_2(u_1,u_2))+f_2(r_1(u_1),u_2)+(-1)^{|u_1|}
f_2(u_1, r_1(u_2))~~.\nn
\eea
In particular, $f_1$ induces a degree zero linear map $f_{1*}$ between the 
cohomologies $H_{r_1}(A)$ and $H_{r'_1}(A')$. 

\paragraph{Definition}
An $A_\infty$ morphism is called a {\em quasi-isomorphism} if
$f_{1*}$ is a degree zero 
isomorphism between the cohomology spaces $H_{r_1}(A)$ and 
$H_{r_1'}(A')$. It is a {\em homotopy} if $f_1$ is a linear isomorphism 
(of degree zero) 
from $A$ to $A'$. It is clear that any homotopy is a quasi-isomorphism.

\subsubsection{The bar construction}

\paragraph{Notation} For any graded vector space $V$, we let $V[1]$
denote its {\em suspension}, i.e. the vector space $V$ endowed with the 
shifted grading $V[1]^k=V^{k+1}$. We let $s:V\rightarrow V[1]$ be the 
identity map of $V$, viewed as a homogeneous morphism of degree $-1$.

If $V$ is a graded vector space, then 
$T(V)=\oplus_{k\geq 1}{V^{\otimes k}}\subset \otimes^* V$ 
denotes its reduced tensor coalgebra, with the (coassociative) 
coproduct $\Delta:T(V)\rightarrow T(V)\otimes T(V)$ defined through:
\be
\Delta(v_1\otimes \dots \otimes v_n)=\sum_{j=1}^{n-1}(v_1\otimes \dots \otimes 
v_j)\otimes (v_{j+1}\otimes \dots \otimes v_n)~~.
\ee
Note that $T(V)$ 
does not contain the summand $V^{\otimes 0}=\C$. In particular, 
it does not have a counit.

\paragraph{Proposition} An $A_\infty$ algebra on the graded vector space $A$ 
is the same as a nilpotent degree one coderivation 
$\partial$ on $T(A[1])$, i.e. a 
homogeneous map of degree one from $T(A[1])$ 
to itself satisfying the conditions:
\be
\Delta\partial=(Id\otimes \partial+\partial\otimes Id)\Delta
\ee
and:
\be
\partial^2=0~~.
\ee

If $\pi:T(A[1])\rightarrow A[1]$ denotes projection on the first component, 
then the product $r_n$ is recovered from the map 
$\partial_n:=\pi\circ \partial|_{A[1]^{\otimes n}}$ by desuspension:
\be
r_n=s^{-1}\circ \partial_n\circ (s^{\otimes n})~~.
\ee

\paragraph{Proposition} Given two $A_\infty$ algebras $A$ and $A'$, 
an $A_\infty$ morphism $f:A\rightarrow A'$ is the same as a coalgebra 
morphism\footnote{This means $\Delta'\circ F=(F\otimes F)\circ \Delta$.} 
$F:T(A[1])\rightarrow T(A'[1])$ which commutes with the 
codifferential $\partial$ and $\partial'$ (i.e. a morphism 
of differential graded coalgebras). The components $f_n$ are obtained 
by desuspension in the obvious manner.

\subsection{$L_\infty$ algebras}
More information on this subject can be found in  
 \cite{Kontsevich_Felder}, \cite{Merkulov_infty} and  
\cite{Lada_Markl} and the references therein. 
The relevance of $L_\infty$ algebras for (closed) string field theory 
was discovered in  \cite{Zwiebach_Witten} and is explained in detail in 
 \cite{Zwiebach_closed}.

\paragraph{Definition} For a graded vector space $V$, 
we let 
$\odot^*V=\oplus_{k\geq 0}{\odot^kV}$ denote the 
associated (graded) symmetric algebra, defined upon dividing 
the free graded 
associative algebra $\otimes^*V=
\oplus_{k\geq 0}{\otimes^kV}$ through the homogeneous ideal 
generated by elements of the form:
\be
u\otimes v-(-1)^{|u||v|}v\otimes u~~,
\ee
where we denote the degree of homogeneous elements $u$ of $V$ by $|u|$. 
Given a permutation $\sigma$ on $n$ elements, we define 
the {\em Koszul sign} $\epsilon(\sigma, u_1\dots u_n)$ through:
\be
u_{\sigma(1)}\odot \dots \odot u_{\sigma(n)}=\epsilon(\sigma; u_1\dots u_n)
u_1\odot \dots \odot u_n~~,
\ee
in the algebra $\odot^* V$.

\paragraph{Definition} Given a graded vector space $V$, we let 
$\Lambda^*V=\oplus_{k\geq 0}{\Lambda^kV}$ denote the 
associated (graded) exterior algebra, defined upon dividing 
the free graded 
associative algebra $\otimes^*V$ through the homogeneous ideal generated
by elements of the form:
\be
u\otimes v+(-1)^{|u||v|}v\otimes u~~.
\ee
If $\sigma$ is a permutation on $n$ elements, we 
define the symbol $\chi(\sigma, u_1\dots u_n)$ through:
\be
u_{\sigma(1)}\wedge \dots \wedge u_{\sigma(n)}=\chi(\sigma; u_1\dots u_n)
u_1\wedge \dots \wedge u_n~~,
\ee
in the exterior algebra $\Lambda^*V$.

\paragraph{Observation} One has:
\be
\chi(\sigma;u_1\dots u_n):=\epsilon(\sigma)\epsilon(\sigma; u_1\dots u_n)~~.
\ee

\paragraph{Definition}
An $L_\infty$ algebra structure on a graded vector space $L$ is 
defined by an infinite sequence of products $m_n:\Lambda^nL\rightarrow L$
which are homogeneous of degree $2-n$ and subject to the constraints:
\be
\label{linf}
\sum_{k+l=n+1}\sum_{\sigma\in Sh(k,n)}{(-1)^{k(l-1)}\chi(\sigma; u_1\dots u_n)
m_l(m_k(u_{\sigma(1)}\dots u_{\sigma(k)}),u_{\sigma(k+1)}\dots u_{\sigma(n)})}=0~~,
\ee
for $n\geq 1$.

The first three conditions in (\ref{linf}) read:
{\footnotesize \bea
\label{l3}
& &m_1^2=0~~\\
& &m_1(m_2(u_1,u_2))=m_2(m_1(u_1),u_2)+(-1)^{|u_1|}m_2(u_1,m_1(u_2))~~\nn\\
& & m_2(m_2(u_1,u_2), u_3))+(-1)^{(|u_1|+|u_2|)|u_3|}
m_2(m_2(u_3, u_1), u_2)+(-1)^{|u_1|(|u_2|+|u_3|)}m_2(m_2(u_2,u_3),u_1)
=\nn\\
& &
-m_1(m_3(u_1,u_2,u_3))-m_3(m_1(u_1), u_2, u_3)-(-1)^{|u_1|}
m_3(u_1, m_1(u_2), u_3)-(-1)^{|u_1|+|u_2|}m_3(u_1, u_2, m_1(u_3))~~.\nn
\eea}\noindent
In particular, the first product $m_1$ is a degree one differential on $L$.
If the triple and higher products all vanish, then an $L_\infty$ algebra 
reduces to a differential graded Lie algebra, with differential given 
by $Q=m_1$ and graded Lie bracket $[u,v]=m_2(u,v)$.

\paragraph{Definition} The cohomology of an $L_\infty$ algebra 
$(L, \{m_n\}_{n\geq 1})$ is the cohomology $H_{m_1}(L)$ of the complex 
$(L, m_1)$.

\subsubsection{Dual description and $L_\infty$ morphisms }

\paragraph{Notation} Given a graded vector space $V$, we define
its {\em reduced symmetric algebra} by 
$S(V)=\oplus_{k\geq 1}{\odot^k V}$ (this is a subspace of 
$\odot^*V$). We view $S(V)$ as 
a (cocommutative, coassociative) 
coalgebra without counit endowed with the coproduct:
\be
\Delta(v_1\odot \dots \odot v_n)=\sum_{j=1}^{n-1}\sum_{\sigma\in Sh(j, n)}
{\epsilon(\sigma,v_1\dots v_n)
(v_{\sigma(1)}\odot \dots \odot v_{\sigma(j)})
\otimes (v_{\sigma(j+1)}\odot \dots \odot v_{\sigma(n)})}~~.
\ee

\paragraph{Proposition} An $L_\infty$ algebra on the graded vector space $L$
is the same as a nilpotent degree one 
coderivation $\delta$ on the coalgebra $S (L[1])$, i.e. a degree one 
linear map from $S(L[1])$ to itself which satisfies:
\be
\Delta\delta=(Id\otimes \delta+\delta\otimes Id)\Delta
\ee
and:
\be
\delta^2=0~~.
\ee

If 
$\pi:S (L[1])\rightarrow L[1]$ is the projection on the first factor, 
then the products $m_n$ are recovered from the maps 
$\delta_n=\pi\circ \delta|_{L[1]^{\odot n}}$ 
by desuspension:
\be
m_n=s^{-1}\circ \delta_n\circ (s^{\odot n})~~,
\ee
where $s^{\odot n}$ is the map induced from $s^{\otimes n}$ by 
graded symmetrization.

\paragraph{Definition} An $L_\infty$ morphism between two $L_\infty$ 
algebras $L$ and $L'$ is specified by a degree zero coalgebra morphism 
$F$ from $S(L[1])$ to $S(L'[1])$ which commutes with the 
codifferentials $\delta$ and $\delta'$. Its components $f_n$ are defined 
by desuspension of $\pi'\circ F|_{L[1]^{\odot n}}$. They are homogeneous 
linear maps of degree $1-n$ from $\Lambda^n L$ to $L'$, which satisfy a 
countable set of constraints equivalent with the condition $\delta'F=F\delta$.

The first of these constraints reads:
\bea
m'_1(f_1(u))&=&m_1(r_1(u))~~.
\eea
Hence $f_1$ induces a degree zero linear map $f_{1*}$ between the 
cohomologies $H_{m_1}(L)$ and $H_{m'_1}(L')$. 

\paragraph{Definition}
An $L_\infty$ morphism is 
a {\em quasi-isomorphism} if
$f_{1*}$ is a (degree zero) 
isomorphism between the cohomology spaces $H_{m_1}(L)$ and 
$H_{m'_1}(L')$. It is a {\em homotopy} if $f_1$ is an isomorphism of graded
vector spaces.

\paragraph{Observation} Since morphisms of codifferential coalgebras 
are closed under composition, we have a natural notion of 
composition of $L_\infty$ morphisms. Hence 
$L_\infty$ algebras and $L_\infty$ morphisms form a category.
An isomorphism in this category is called an $L_\infty$ {\em isomorphism}.
It can be shown  \cite{Kontsevich_Felder} that an $L_\infty$-morphism 
$f:L\rightarrow L'$ 
is an isomorphism if and only if it is a homotopy.

\paragraph{Theorem  \cite{Kontsevich_Felder}} 

Every $L_\infty$ quasi-isomorphism  
$(L, \{m_n\}_{n\geq 1})\stackrel{f}{\rightarrow}(L', \{m'_n\}_{n\geq 1})$
admits a {\em quasi-inverse} i.e. there exists an $L_\infty$ quasi-isomorphism 
$(L', \{m'_n\}_{n\geq 1})\stackrel{g}{\rightarrow}(L, \{m_n\}_{n\geq 1})$
such that $g_{1*}=(f_{1*})^{-1}$.

\subsubsection{Morphisms to a dG Lie algebra}

This subsection gives the explicit conditions satisfied by an $L_\infty$ 
morphism from an $L_\infty$ algebra to a dG Lie algebra. This is the case 
relevant for Section 3 of the paper.

\paragraph{Proposition  \cite{Lada_Markl}}
Given an $L_\infty$ algebra 
$(L, \{m_n\}_{n\geq 1})$ and a dG Lie algebra 
$(L', Q, [.,.])$, an $L_\infty$ 
{\em morphism} $f=\{f_n\}_{n\geq 1}$ from $L$ to $L'$ amounts to 
a collection of 
maps $f_n:\Lambda^nL\rightarrow L'$ which are homogeneous of degree $1-n$ 
and satisfy the conditions:
\bea
\!\!\!\!\!\!\!\!\!\!\!\!
& &Q f_n(u_1,\dots,u_n)+
\sum_{j+k=n+1}\sum_{\sigma\in Sh(k,n)}{(-1)^{k(j-1)+1}\chi(\sigma)
f_j(m_k(u_{\sigma(1)},\dots ,u_{\sigma(k)}),u_{\sigma(k+1)},\dots ,u_{\sigma(n)})}~~~~~~~~~~~~~~~~~~~~~\nn\\
& &=\sum_{s+t=n}
\sum_{\tiny\begin{array}{c}\tau\in Sh(s,n)\\\tau(1)<\tau(s+1)\end{array}}
{(-1)^{s+(t-1)\sum_{p=1}^s{|u_{\tau(p)}|}}
\chi(\tau)[f_s(u_{\tau(1)},\dots ,u_{\tau(s)}),
f_t(u_{\tau(s+1)},\dots ,u_{\tau(n)})]}~~.~~~~~~~~~
\eea
for $n\geq 1$.

\subsubsection{The deformation functor of an $L_\infty$ algebra}

\paragraph{Definition  \cite{Kontsevich_Felder, Merkulov_infty}} 
Given an $L_\infty$ algebra $(L,\{m_n\})$, 
its (unextended) {\em deformation functor} $Def^0_L$ 
is the functor from the category of local Artin algebras to the category of 
sets which associates to the algebra $B$ the moduli space:
\be
Def_L(B)=\{\phi\in (L\otimes m_B)^1|
\sum_{n\geq 1}{\frac{(-1)^{n(n+1)/2}}{n!}m_n(\phi^{\otimes n})}=0\}
/{\cal G}_L(B)~~
\ee
of solutions to the `homotopy Maurer-Cartan equation' taken modulo the 
action ${\cal G}_L(B)$ generated by infinitesimal transformations of the form:
\be
\label{infty_gauge}
\phi\rightarrow \phi'=\phi-\sum_{n\geq 1}{\frac{(-1)^{n(n-1)/2}}{(n-1)!}
m_n(\alpha\otimes \phi^{\otimes n-1})}~~,
\ee
where $\alpha\in (L\otimes m_B)^0$. Here $m_B$ is the maximal ideal of $B$.

\paragraph{Observations} 1. The technical device of tensoring with an Artin 
algebra leads to {\em formal deformations}, thereby eliminating problems 
of convergence. The associated formal moduli space is obtained by 
representing the deformation functor.

2. The fact that infinitesimal gauge transformations of the form 
(\ref{infty_gauge}) preserve the homotopy Maurer-Cartan equation is checked 
by direct computation in Appendix B. The explicit form of these 
transformations differs from that given in  \cite{Merkulov_infty}  
through the sign factors.

\paragraph{Theorem \cite{Kontsevich_Felder, GM}}

If two $L_\infty$ algebras $L$ and $L'$ are quasi-isomorphic, then their 
deformation functors are equivalent, $Def^0_L\approx Def^0_{L'}$.

\subsection{Commutator algebra of an $A_\infty$ algebra}

\paragraph{Definition \cite{Lada_Markl}}

Given an $A_\infty$ algebra $(A, \{r_n\})$, we define new products $m_n$ 
by graded antisymmetrization of $r_n$:
\be
\label{cm}
m_n(u_1\dots u_n)=\sum_{\sigma\in S_n}{\chi(\sigma, u_1\dots u_n)r_n(u_{\sigma(1)}\dots 
u_{\sigma(n)})}~~.
\ee
It is then easy to check that $(A, \{m_n\})$ is an $L_\infty$ algebra, 
called the {\em commutator algebra} of $(A, \{r_n\})$. 
If the higher products $r_n$ $(n\geq 3)$ vanish, so that $A$ is a differential
graded associative algebra, then its commutator algebra is simply the 
associated differential graded Lie algebra.

\paragraph{Dual description \cite{Lada_Markl}}
Let $S$ be the injective coalgebra morphism from $S(A[1])$ to 
$T(A[1])$ defined through:
\be
S(w_1\odot ~\dots ~\odot w_n)=
\sum_{\sigma\in S_n}{\epsilon(\sigma, w_1~\dots ~w_n)
w_\sigma(1)\otimes\dots \otimes w_{\sigma(n)}}~~.
\ee
This map allows us to view $S(A[1])$ as a sub-coalgebra of $T(A[1])$.
One then has the following:

\paragraph{Proposition \cite{Lada_Markl}}
The $A_\infty$ codifferential $\partial$ preserves the subspace $ImS$
of $T(A[1])$. In fact, there exists a unique degree one codifferential 
$\delta$ on $S(A[1])$ such that:
\be
\partial S=S\delta~~.
\ee
Moreover, the codifferential $\delta$ defines the commutator 
$L_\infty$ structure (\ref{cm}).

\paragraph{Proposition}

Given an $A_\infty$ morphism $f=\{f_n\}_{n\geq 1}$ between two 
$A_\infty$ algebras $(A, \{r_n\})$ and $(A', \{r'_n\})$,  
we define its (graded) antisymmetrization $g=\{g_n\}_{n\geq 1}$ through:
\be
\label{gs}
g_n(u_1\dots u_n)=\sum_{\sigma \in S_n}{\chi(\sigma, u_1\dots u_n)
f_n(u_{\sigma(1)}\dots u_{\sigma(n)})}~~.
\ee
Then $g$ is an $L_\infty$ morphism between the commutator $L_\infty$ algebras 
$(A,\{m_n\}_{n\geq 1})$ and $(A',\{m'_n\}_{n\geq 1})$. Moreover, if 
$f$ is a quasi-isomorphism, then so is $g$.

\paragraph{Proof} Consider the associated 
map $F:T(A[1])\rightarrow T(A'[1])$, which satisfies $\partial'F=F\partial$.
It is easy to see that (\ref{gs}) correspond to the unique map 
$G:S(A[1])\rightarrow S(A'[1])$ with the property:
\be
S'G=FS~~.
\ee

We have to show that $\delta' G=G\delta$. In view of injectivity of $S'$, 
it suffices to show that $S'\delta'G=S'G\delta$. This follows from the 
chain of equalities:
\bea
S'\delta'G=\partial'S'G=\partial'FS=F\partial S=FS\delta=S'G\delta~~.
\eea
If $f$ is a quasi-isomorphism, then $f_1$ induces an isomorphism on 
cohomology. It is clear from (\ref{gs}) that $g_1=f_1$, hence $g$ 
is a quasi-isomorphism as well.

\section{Infinitesimal gauge transformations in an $L_\infty$ algebra}

In this appendix we give a direct proof of the fact that transformations
(\ref{infty_gauge}) preserve the homotopy Maurer-Cartan equation 
(\ref{mc_L}). The explicit form of these transformations is of independent 
interest, for example for a better understanding of 
moduli spaces in closed string field theory \cite{Zwiebach_closed}.
Let us consider a degree one solution $\phi$ of the homotopy Maurer-Cartan 
equations:
\be
\label{eq1}
\sum_{l\geq 1}{\frac{(-1)^{l(l+1)/2}}{l!}m_l(\phi^{\otimes l})}=0~~,
\ee
and an infinitesimal gauge variation of the type (\ref{infty_gauge}):
\be
\label{eq2}
\delta\phi=-\sum_{k\geq 1}{\frac{(-1)^{k(k-1)/2}}{(k-1)!}m_k(\alpha\otimes 
\phi^{\otimes k-1})}~~,
\ee
where $\alpha$ is a degree zero element of $L$.
Then $\phi+\delta\phi$ satisfies the homotopy Maurer-Cartan equation
to first order in $\alpha$ provided that the variation $\delta M $ 
of the left hand side of (\ref{eq1}) vanishes. This first order variation 
is given by: 
\be
\delta M=\sum_{l\geq 1}{\frac{(-1)^{l(l+1)/2}}{(l-1)!}
m_l(\delta\phi, \phi^{\otimes l-1})}~~.
\ee
Upon substituting (\ref{eq2}), this becomes:
\be
\label{eq3}
\delta M= -\sum_{k,l\geq 1}{\frac{(-1)^{k(k-1)/2+l(l+1)/2}}{(k-1)!(l-1)!}
m_l(m_k(\alpha\otimes \phi^{\otimes k-1}), \phi^{\otimes l-1})}~~.
\ee
We will show that vanishing of this quantity follows from the 
homotopy Lie identities (\ref{L_inf}). For this, we start by re-writing the 
latter in the form:
\be
\label{eq4}
\sum_{k+l=n+1}\sum_{\sigma\in Sh(k,n)}{(-1)^{l(k-1)}\chi(\sigma; u_1\dots u_n)
m_l(m_k(u_{\sigma(1)}\dots u_{\sigma(k)}),u_{\sigma(k+1)}\dots u_{\sigma(n)})}=0~~,
\ee
which is obtained upon multiplying (\ref{L_inf}) with $(-1)^{n+1}=(-1)^{k+l}$.
This allows us to trade the sign factor $(-1)^{k(l-1)}$ in (\ref{L_inf}) for 
$(-1)^{l(k-1)}=(-1)^{k(l-1)}(-1)^{n+1}$, a trick which will be useful later. 

We now apply (\ref{eq4}) to the elements $u_1=\alpha$ and $u_2=\dots =u_n=\phi$.
We want to extract the explicit form of the unsigned and signed summands 
$s(k,n)=m_l(m_k(u_{\sigma(1)}\dots u_{\sigma(k)}),u_{\sigma(k+1)}
\dots u_{\sigma(n)})$
and $S(k,n)=\chi(\sigma; u_1\dots u_n)s(k,n)$ for these values of 
$u_j$. For this, note that each $(k,n)$-shuffle $\sigma$ 
determines a partition 
$\{1\dots n\}=I\sqcup J$ of the set $\{1\dots n\}$ through:
\be
I=\{\sigma(1)~\dots ~\sigma(k)\}~~,~~J=\{\sigma(k+1)~\dots ~\sigma(n)\}~~.
\ee
Since $1$ must belong to precisely one of these two sets, and since 
$\sigma(1)<\dots <\sigma(k)$ and $\sigma(k+1)<\dots <\sigma(n)$, we can divide the 
set of $(k,n)$-shuffles into two disjoint subsets: 

(1) The set $Sh_{+}(k,n)$, of shuffles 
for which $1$ belongs to $I$, i.e. $\sigma(1)=1$

(2) The set $Sh_{-}(k,n)$, of shuffles for which $1$ belongs to $J$, i.e. 
$\sigma(k+1)=1$. 

It is clear that $Sh_{+}(k,n)$ contains $C_{n-1}^{k-1}$ elements, 
while $Sh_{-}(k,n)$ has cardinality $C_{n-1}^{k}$ ($C_{a}^b$ denotes 
the number of ways of choosing $b$ out of $a$ elements). Moreover, if 
the shuffle $\sigma$ belongs to $Sh_{+}(k,n)$, then it easy to check that:
\be
s(k,n)=m_l(m_k(\alpha,\phi^{\otimes k-1}),~\phi^{\otimes l-1})~~{\rm and}~~
S(k,n)=s(k,n)~~,
\ee 
while if $\sigma$ belongs to $Sh_{-}(k,n)$ then one has:
\be
s(k,n)=(-1)^l~m_l(m_k(\phi^{\otimes k}),~\phi^{\otimes l-2},\alpha)~~
{\rm and}~~S(k,n)=(-1)^k~s(k,n)~~.
\ee
Upon substituting this in (\ref{eq4}) and multiplying everything by 
$\frac{(-1)^{n(n+1)/2}}{(n-1)!}$, we obtain:
\be
\label{eq_sum}
\Sigma_{+}(n)+\Sigma_{-}(n)=0~~,
\ee
where:
\be
\label{eq5}
\Sigma_{+}(n)=\sum_{k+l=n+1}{
\frac{(-1)^{k(k-1)/2+l(l+1)/2}}
{(k-1)!(l-1)!}m_l(m_k(\alpha,\phi^{\otimes k-1}),
\phi^{\otimes l-1})}~~.
\ee
and
\be
\label{eq6}
\Sigma_{-}(n)=\sum_{k+l=n+1}{
\frac{(-1)^{k(k+1)/2+l(l-1)/2}}{k!(l-2)!}m_l(m_k(\phi^{\otimes k}),
\phi^{\otimes l-2},\alpha)}~~.~~~
\ee
To arrive at these expressions, we used the identities:
\be
\frac{k(k-1)}{2}+\frac{l(l+1)}{2}+l(k-1)=\frac{n(n+1)}{2}~{\rm~and~}
\frac{k(k+1)}{2}+\frac{l(l-1)}{2}+k(l-1)=\frac{n(n+1)}{2}~~~~
\ee
in order to re-write the various sign factors. Comparing eqs. (\ref{eq5}) and 
(\ref{eq3}) shows that:
\be
\sum_{n\geq 1}{\Sigma_{+}(n)}=-\delta M~~.
\ee
On the other hand, the homotopy Maurer-Cartan equation (\ref{eq1})
implies:
\be
\sum_{n\geq 1}{\Sigma_{-}(n)}=0~~.
\ee
Hence summing the identities (\ref{eq_sum}) over $n$ gives the desired 
conclusion:
\be
\delta M=0~~.
\ee


\begin{thebibliography}{100}
\bibitem{Douglas_quintic}{ 
I.~Brunner, M.~R.~Douglas, A.~Lawrence, C.~Rom
elsberger
{\em D-branes on the Quintic}, JHEP {\bf 0008} (2000) 015, hep-th/9906200.}
\bibitem{Kachru1}{ 
S.~Kachru, S.~Katz, A.~Lawrence, J.~McGreevy, 
{\em Open string instantons and superpotentials }, 
Phys.Rev. {\bf D62} (2000) 026001, hep-th/9912151.}
\bibitem{com1}{C.~I.~Lazaroiu, 
{\em Generalized complexes and string field theory}, JHEP {\bf 06} (2001) 052, 
hep-th/0102122.}
\bibitem{com3}{C.~I.~Lazaroiu, {\em 
Unitarity, D-brane dynamics and D-brane categories}, hep-th/0102183. }
\bibitem{Diaconescu}{D.~E.~Diaconescu,
{\em Enhanced D-Brane Categories from String Field Theory}, 
hep-th/0104200, JHEP {\bf 06} (2001) 016.}
\bibitem{sc}{C.~I.~Lazaroiu, {\em Graded Lagrangians, exotic topological 
D-branes  and enhanced triangulated categories}, JHEP {\bf 06} (2001)064, 
hep-th/0105063.}
\bibitem{bv}{C.~I.~Lazaroiu, R.~Roiban and D.~Vaman, 
{\em Graded Chern-Simons field theory and graded topological D-branes}, 
hep-th/0107063.}
\bibitem{Kontsevich}{M.~Kontsevich, {\em Homological algebra of mirror 
symmetry},  Proceedings of the International 
Congress of Mathematicians (Zurich, 1994), 120--139, Birkhauser, 
alg-geom/9411018.}
\bibitem{Kontsevich_Felder}{M.~Kontsevich, 
{\em Deformation quantization of Poisson Manifolds}, I, 
mat/9709010.}
\bibitem{Kontsevich_Soibelman}{M.~Kontsevich, Y.~Soibelman, 
{\em Homological mirror symmetry and torus fibrations}, math.SG/0011041.}
\bibitem{Merkulov_defs}{S.~A.~Merkulov, {\em $L_{\infty}$-algebra of an 
unobstructed deformation functor}, math.AG/9907031 }
\bibitem{Merkulov_infty}{S.~A.~Merkulov, 
{\em  Frobenius infinity invariants of homotopy Gerstenhaber algebras I},
math.AG/0001007.}
\bibitem{Kapranov_derived_defs}{ I.~Ciocan-Fontanine, M.~Kapranov, 
{\em  Derived Quot schemes},  math.AG/9905174.}
\bibitem{Fukaya}{K.~Fukaya, 
{\em  Morse homotopy,  $A^\infty$-category and  Floer  homologies}, in
{\em  Proceedings of  the  GARC Workshop  on  Geometry and  Topology},
ed. by H.~J.~Kim, Seoul  national University (1994), 1-102; {\em Floer
homology, $A^\infty$-categories and topological field theory}, in {\em
Geometry and Physics}, Lecture  Notes in Pure and Applied Mathematics,
{\bf 184},  pp 9-32, Dekker, New  York, 1997; {\em  Floer homology and
Mirror       symmetry,      I},       preprint       available      at
$http://www.kusm.kyoto-u.ac.jp/~\tilde{}~fukaya/fukaya.html.$}
\bibitem{Fukaya2}{K. Fukaya, Y.-G. Oh, H.~Ohta, K.~Ono, 
{\em Lagrangian intersection Floer theory - anomaly and obstruction}, 
  preprint       available      at
$http://www.kusm.kyoto-u.ac.jp/~\tilde{}~fukaya/fukaya.html.$}
\bibitem{boundary}{C.~I.~Lazaroiu, 
{\em  Instanton amplitudes in open-closed topological string theory}, 
 hep-th/0011257.}
\bibitem{Polishchuk}{A.~Polishchuk, 
{\em Homological mirror symmetry with higher products}, math.AG/9901025}
\bibitem{Witten_SFT}{E.~Witten, {\em Noncommutative geometry and string 
field theory}, Nucl. Phys, {\bf B268} (1986) 253.}
\bibitem{Thorn}{C.~B.~Thorn, {\em String field theory},~Phys.~Rept. 
{\bf175} (1989) 1.}
\bibitem{AS}{S.~Axelrod, I~.M.~Singer, {\em Chern-Simons perturbation 
theory}, in Proceedings of the XXth International Conference 
on Differential Geometric Methods in Theoretical Physics, World Sci. 
Publishing, River Edge, NJ, 1992,  hep-th/9110056.}
\bibitem{Zwiebach_closed}{B.~Zwiebach,
{\em Closed string field theory: Quantum action and the B-V master equation},
Nucl.~Phys. {\bf B390} (1993) 33, hep-th/9206084.}
\bibitem{Merkulov}{ S.A. Merkulov,
{\em Strong homotopy algebras of a Kahler manifold}, math.AG/9809172} 
\bibitem{DVV}{R.~Dijkgraaf, H.~Verlinde, E.~Verlinde,
{\em  Topological strings in $d<1$},
Nucl.Phys. {\bf B352} (1991) 59-86 }
\bibitem{Hofman1}{C.~Hofman, W.~K.~Ma, 
{\em  Deformations of Topological Open Strings},  JHEP {\bf 0101} (2001) 035.}
\bibitem{Witten_CS}{
E.~Witten, {\em Chern-Simons gauge theory as a string theory}, 
The Floer memorial volume, 637--678, Progr. Math. {\bf 133}, Birkhauser, Basel,
1995, hep-th/9207094.}
\bibitem{Lada_Markl}{T.~Lada, M.~Markl, 
{\em Strongly homotopy Lie algebras}, Comm. Algebra {\bf 23} (1995), 
no. 6, 2147--2161, hep-th/9406095. }
\bibitem{Keller}{B.~Keller, 
{\em Introduction to A-infinity algebras and modules}, 
math.RA/9910179 .}
\bibitem{Aspinwall}{P.~S.~Aspinwall, A.~Lawrence, 
{\em Derived Categories and Zero-Brane Stability}, 
hep-th/0104147.}
\bibitem{bcft_defs}{A.~Recknagel, V.~Schomerus, 
{\em Boundary Deformation Theory and Moduli Spaces of D-Branes},
Nucl.Phys. {\bf B545} (1999) 233-282
hep-th/9811237.}
\bibitem{Douglas_Kontsevich}{M.~Douglas,
{\em D-branes, Categories and N=1 Supersymmetry}, hep-th/0011017.}
\bibitem{McLean}{R.~C.~Mc Lean, 
{\em Deformations of calibrated submanifolds}, 
Comm. Anal. Geom {\bf 6}, 1998, 705-747.}
\bibitem{Hitchin}{N.~Hitchin, 
{\em  Lectures on Special Lagrangian Submanifolds},
 math.DG/9907034.}
\bibitem{Zaslow_recent}{N.~C.~Leung, S.-T.~Yau, E.~Zaslow, 
{\em From Special Lagrangian to Hermitian-Yang-Mills via Fourier-Mukai 
Transform}, math.DG/0005118.}
\bibitem{Merkulov_sl}{S.~A.~Merkulov, {\em The extended moduli space of special Lagrangian submanifolds}, Commun.Math.Phys. {\bf 209} (2000) 13-27, 
math.AG/9806083.}
\bibitem{Gaberdiel}{M.~Gaberdiel, B.~Zwiebach, {\em 
Tensor constructions of open string theories I:Foundations,
} Nucl. Phys {\bf B505} (1997) 569, hep-th/9705038.}
\bibitem{Zwiebach_open}{
B.~Zwiebach, {\em Oriented open-closed string theory revisited}, 
Annals. Phys. {\bf 267} (1988), 193, hep-th/9705241.}
\bibitem{Katz}{S.~Katz, {\em 
Versal deformations and superpotentials for rational curves in smooth 
threefolds},  math.AG/0010289.}
\bibitem{top}{C.~I.~Lazaroiu, 
{\em On the structure of open-closed topological field 
theory in two dimensions}, Nucl.Phys. {\bf B603} (2001) 497-530,
hep-th/0010269.}
\bibitem{Polishchuk_ell2}{A.~Polishchuk, 
$A_{\infty}$-structures on an elliptic curve, math.AG/0001048~~. }
\bibitem{Zwiebach_Witten}{B.~Zwiebach, E.~Witten, 
{\em Algebraic structures and differential geometry in 2-d string 
theory}, Nucl.Phys. {\bf B377}(1992)55-112. }
\bibitem{GM}{W.~M.~Goldman, J.~J.~Millson, 
{\em The homotopy invariance of the Kuranishi space}, Illinois J. Math.
{\bf 34} (1990), no. 2, 337--367. }
\bibitem{Proute}{A.~Proute, 
{\em Algebres differentielles fortement homotopiquement associatives}, 
These d'Etat, Univ. Paris VII, 1984.}
\bibitem{Sullivan}{M.~Chas, D.~Sullivan, {\em String Topology}, 
math.GT/9911159.}
\bibitem{Hofman2}{C.~Hofman, W.~K.~Ma,
{\em Deformations of Closed Strings and Topological Open Membranes}, 
JHEP {\bf 0106} (2001) 033, hep-th/0102201.}
\bibitem{JS}{J.~S.~Park, {\em Topological open p-branes}, 
hep-th/0012141.}
\bibitem{GH}{P.~Griffiths and J.~Harris, {\em Principles of algebraic
      geometry}, Wiley, 1994. }
\end{thebibliography}
\end{document}